\documentclass[%
 reprint,
 amsmath,amssymb,
 aps,
]{revtex4-2}

\usepackage{graphicx}
\usepackage{dcolumn}
\usepackage{bm}
\usepackage{hyperref}
\usepackage{cleveref}
\crefname{equation}{Eq.}{Eqs.}
\Crefname{equation}{Equation}{Equations}
\crefname{figure}{Fig.}{Figs.}
\Crefname{figure}{Figure}{Figures}
\crefname{section}{Sect.}{Sects.}
\Crefname{section}{Section}{Sections}

\usepackage{tikz}
\usepackage{tikz-cd}
\usepackage{lipsum}
\usepackage{amsthm}

\newcommand{\pp}[1]{\left({#1}\right)}
\newcommand{\bb}[1]{\left[{#1}\right]}

\usepackage{physics}
\usepackage{centernot}

\usepackage{mathtools}

\newcommand{\mysim}[1]{\stackrel{\mathclap{\normalfont\mbox{#1}}}{\preceq}}
\newcommand{\mynsim}[0]{\npreceq}

\DeclareFontFamily{U}{stixextrai}{}
\DeclareFontShape{U}{stixextrai}{m}{n}
 { <-> stix-mathtt }{}

\newcommand{\lltriangle}{{\usefont{U}{stixextrai}{m}{n}\symbol{153}}}

\makeatletter
\providecommand{\vrectangle}{\mkern1mu\mathpalette\v@rectangle\relax\mkern1mu}
\newcommand{\v@rectangle}[2]{%
  \hbox{
  \fboxrule=0.5\fontdimen 8
    \ifx#1\displaystyle\textfont\else
    \ifx#1\textstyle\textfont\else
    \ifx#1\scriptstyle\scriptfont\else
    \scriptscriptfont\fi\fi\fi 3
  \fboxsep=-\fboxrule
  \fbox{$\m@th#1\phantom{(}$}%
  }
}
\makeatother

\usepackage{nicematrix}

\begin{document}

\preprint{APS/123-QED}

\title{Dynamical GKP codes}

\author{Jonathan Pelletier}

\author{Baptiste Royer}%
\affiliation{Département de Physique and Institut Quantique, Université de Sherbrooke, Sherbrooke, Qc. J1K 2R1, Canada}

\date{\today}

\begin{abstract}
Dynamical codes protect quantum information with sequences of measurements, generalizing the stabilizer formalism of static codes. However, the dynamical formalism has been developed exclusively for qubit systems. While Gottesman–Kitaev–Preskill (GKP) codes are promising candidates for quantum computing that encode qubits in the continuous phase space of oscillators, no dynamical formalism has been developed for them. In this work, we bridge this gap and introduce dynamical GKP codes, combining continuous-variable lattice encodings with time-dependent translation measurement dynamics. To obtain the update rules and the distance of a given measurement sequence, we first generalize GKP codes to subsystem GKP codes using scaled symplectic lattices. Building on this structure, we adapt the ancestry, detector, and benign-error machinery originally developed for dynamical qubit codes. From this construction, we obtain a general and efficient method to compute the syndrome and distance of a dynamical GKP code from a persistent stabilizer lattice and a bounded inference window. We illustrate the framework with the example of the two-mode square-hexagonal Floquet GKP code. This code has reduced measurement weight compared to both the square and hexagonal GKP codes, while combining their set of logical gates, resulting in easier implementation of the full single-qubit Clifford group. Finally, we show that measurements with irrational symplectic overlap, which have no analog in discrete qubit codes, reduce to destructive homodyne detection in the single-mode case. These results extend an important qubit-code tool to continuous-variable systems and identify a new bridge between discrete stabilizer measurements and continuous homodyne detection.
\end{abstract}

\maketitle

Quantum computers are only as useful as their ability to protect quantum information from noise. In standard, qubit-based quantum error correction (QEC) protocols, information is spread redundantly across many two-level systems so that local errors can be detected and corrected without destroying the encoded state. An alternative strategy is to encode the information in the bosonic modes of harmonic oscillators, such as electromagnetic cavities or mechanical resonators. The Gottesman–Kitaev–Preskill (GKP) code is one leading example of this continuous-variable (CV) approach, and while it is designed to protect against small, random displacements, this code also partially corrects the main error channel that afflicts oscillators: loss \cite{albert_2018,noh_2019a}. The GKP encoding has now been realized experimentally in superconducting circuits \cite{sivak_2023a,campagne-ibarcq_2020}, trapped ions \cite{fluhmann_2019} and optical systems \cite{larsen_2025a}. Because a single well-protected oscillator can outperform many physical qubits, GKP codes are considered one of the most hardware-efficient routes toward fault-tolerant quantum computation. Understanding how to extend and combine them with other QEC techniques is of direct interest to experimentalists and theorists working across quantum optics, superconducting circuits and trapped ions.

A substantial fraction of recent progress in bosonic quantum error correction has come from a translation exercise: taking tools and results developed for qubit quantum error correction and adapting them to the CV setting of GKP codes. This strategy has been successful, in large part because GKP codes are built from a discrete structure, a lattice, inside a continuous one. The lattices in phase space reproduce the algebraic structure of qubit stabilizer codes, which allows to carry over many qubit-code techniques to the world of lattice encoding. This adaptation process also highlights core differences between the two settings, and it is often precisely in these differences that GKP codes reveal their advantages. For example, CV systems allow for different types and sizes of encoding without modification to the devices, they support gate implementations that have no direct qubit analog, and they open the door to measurement strategies, such as homodyne detection, that are native to bosonic systems.

Among the qubit-code techniques that have not yet been translated to the bosonic setting are dynamical and Floquet codes. The formalism of dynamical codes is a generalization of the stabilizer formalism that fully incorporate the notion of measurements with QEC. In particular, it includes the description of all stabilizer codes. Dynamical codes have attracted considerable attention, in part because they achieve strong error protection using only sequences of low-weight measurements rather than the high-weight stabilizer measurements typically required by static codes. This is especially appealing for near-term hardware, where the fidelity of a stabilizer measurement generally degrades with its weight \cite{paetznick_2023}. GKP codes share exactly this sensitivity: the fidelity of a GKP stabilizer measurement is known to degrade with the size of the displacement being measured \cite{royer_2020a}. This suggests that a dynamical code built from a sequence of short-displacement measurements could offer the same hardware advantages for bosonic hardware that dynamical qubit codes offer for qubit hardware. Besides advantageous stabilizer measurements, dynamical codes naturally describe the code switching mechanism, which is an important tool to combine interesting sets of logical gates \cite{anderson_2014}. Using code switching between specific static GKP codes, one can combine their set of logical gates and achieve a complete and high-fidelity set of logical gates. At the same time, some GKP hardware provides native access to homodyne detection \cite{bourassa_2021a}, that have no natural counterpart in the discrete Pauli measurements of dynamical qubit codes. Understanding how such measurements fit into a dynamical framework is therefore not just a technical convenience but an opportunity to uncover new physics at the interface of discrete and continuous encodings.

Many works have been done on dynamical qubit codes, and a general theory has since been developed to compute their distance by tracking how spacetime errors propagate across measurement rounds. This theory relies on two central ingredients: the ancestry of a stabilizer, which traces how a given measurement outcome depends on the outcomes of previous rounds, and the classification of undetectable errors into equivalence classes using a set of benign error generators. Bringing this machinery to GKP codes, however, is not a direct transcription. Qubit stabilizer groups are finite and every Pauli operator squares to the identity, which guarantees that the size of the instantaneous stabilizer group (ISG) saturates in a bounded number of rounds. This guarantee disappears a priori for a lattice of translation operators, whose powers are generally nontrivial stabilizers rather than the identity. Measurement outcomes for GKP codes are also modular phases rather than binary values. A further complication, with no counterpart in the qubit case, arises when a measured translation has an irrational symplectic overlap with the existing stabilizer lattice. Such a measurement annihilates the lattice structure and instead behaves as a continuous, homodyne-like measurement of the oscillators.

In this paper, we develop a lattice-based framework for dynamical and subsystem GKP codes that addresses the differences with their qubit counterpart. The paper is organized as follows. In \cref{sec:Prelim}, we review the key definitions and notation for dynamical qubit codes, subsystem codes and GKP codes. In \cref{sec:SubGKP}, we formulate subsystem GKP codes as scaled versions of a usual lattice encoding, which lets us define the bare, dressed, and gauge lattices, as one would get from a subsystem qubit code. We also derive a corresponding notion of subsystem distance, in direct analogy with qudit subsystem codes. In \cref{sec:DynGKP}, we define dynamical GKP codes as sequences of translation measurements and derive explicit update rules for the stabilizer lattice, its phase vector, and the logical representatives after each round. In \cref{sec:EC_DGKP}, using the update rules, we adapt the ancestry and the benign-error constructions of dynamical qubit codes to the continuous setting. These tools, combined with the identification of a persistent lattice and a bounded inference window, yield a general method for computing the syndrome and the distance of a dynamical GKP code. Those tools are illustrated with the example of the periodic square-hexagonal GKP code. This example exhibits the expected advantages of dynamical codes: the reduced weight measurements and the combination of the gate set from the square and the hexagonal GKP code. In \cref{subsec:hom_meas}, we examine the case of measurements with irrational symplectic overlap and show that they reduce to homodyne-type detection on the encoded oscillators, establishing a precise connection between a class of continuous measurements native to bosonic hardware and the discrete, lattice-based framework used throughout the rest of the paper. We conclude in \cref{sec:Conclusion}.
\section{\label{sec:Prelim} Preliminaries and notation}
In this paper, we merge the construction of dynamical codes with Gottesman–Kitaev–Preskill codes, requiring extensive knowledge of both of them. We begin by setting key definitions and notations that are necessary for the following development. 

\subsection{\label{subsec:Prel_Dyn} Dynamical qubit codes}
In recent years, extensive work has been done to understand dynamical qubit codes and their performance \cite{hastings_2021,delfosse_2023,fu_2025}. The series of works was completed by \cite{blackwell_2025}, where a definitive answer on the question of the distance of dynamical qudit codes was given. In the current section, we recall some of their results.

Consider the codespace stabilized by a commutative subgroup of the Pauli group $\mathcal{P}_t^n$, called the instantaneous stabilizer group ISG$(t)$ at time $t$. Following a Pauli measurement of $M_{t+1}$ in the next round, only the elements of $\text{ISG}\pp{t}$ that commute with $M_{t+1}$ can have a value consistent with the new measurement. Therefore, the updated instantaneous stabilizer group is given by 
\begin{equation}
  \text{ISG}\pp{t+1} = \left< C_{\text{ISG}\pp{t}}\pp{M_{t+1}}, M_{t+1} \right>,
\end{equation}
where $C_{\text{ISG}\pp{t}}\pp{M_{t+1}}$ denotes the centralizer of $M_{t+1}$ inside $\text{ISG}\pp{t}$. When $M_{t+1}\in \text{ISG}\pp{t}$, the outcome of the measurement is already fixed and the instantaneous stabilizer group is unchanged $\text{ISG}\pp{t}=\text{ISG}\pp{t+1}$. Otherwise, when $M_{t+1}\notin \text{ISG}\pp{t}$, the outcome is random.

A dynamical code is defined by a time sequence $\left\{\bm{M}_t\right\}_{t\geq 1} $ of commuting Pauli measurements $\bm{M}_t = \left\{M_{t,i} \right\}$ such that $\bb{M_{t,i},M_{t,j}}=0$. If the time sequence $\bm{M}_t$ is periodic, the dynamical code is also called a Floquet code. While static stabilizer codes protect the logical information from small weight Pauli errors, dynamical codes protect against time sequences of Pauli errors of small weight. We denote the set of time sequences of Pauli errors, also called a spacetime Pauli error, by $\mathcal{E}$ :
\begin{equation}
  E=\pp{...,E_t,...}\in \mathcal{E}= \bigoplus_{t\geq0} \mathcal{P}_t^n,
\end{equation} 
with natural product rule $E\tilde{E}=\pp{...,E_t\tilde{E}_t,...}$. A spacetime error $E$ indicate that an error $E_t$ has occurred in-between the measurement round $\bm{M}_t$ and $\bm{M}_{t+1}$. The weight of a spacetime error is the sum of the weight of instantaneous errors. We will often use the notation $E_t$ to mean both the Pauli operator in $\mathcal{P}_{t}^{n}$ and its spacetime generalization 
\begin{equation}
  \pp{...,I_{t-1},E_t,I_{t+1},...}.
\end{equation}
While a good distance of each ISG composing a dynamical code is necessary to obtain a good dynamical distance, it is not a guarantee. It is possible for a low weight spacetime error to grow with the measurement schedule and result in a logical failure at a latter time. To compute the dynamical distance of a code, it is crucial to find all undetectable errors while also being able to tell which spacetime error will lead to a logical failure in the future. The work from Blackwell \textit{et al.} \cite{blackwell_2025} introduced the machinery to compute the dynamical distance of a dynamical code. Their work can be split into two parts. The first part concerns the construction of the detectors using the ancestry of stabilizers and the second separates spacetime errors into equivalence classes using benign errors. Both are tied together using the inference window. In the remainder of the present section, we summarize their results.

Given a stabilizer $S_t$ in $\text{ISG}\pp{t}$ and a measurement sequence $\left\{M_t\right\}$ with outcomes $\left\{m_t\right\}$, such that $m_t^{*} M_t \,\in\, \text{ISG}\pp{t} $, the ancestry of $S_t$ at time $\tau \leq t$ is the spacetime Pauli operator that describes how $S_t$ was built through the rounds of measurement. Precisely, the ancestry is defined in a recursive fashion as
\begin{widetext}
\begin{equation}
  \text{ance}_{\tau}^{t}\pp{S_t} = \begin{cases}
I_t\cdot \text{ance}_{\tau}^{t-1}\pp{\bb{S_t}_{t-1}}  & \text{if } \bb{S_t}_{t-1}\in\text{ISG}\pp{t-1}\quad \& \quad t> \tau,\\
m_t^{\dagger}M_t \cdot \text{ance}_{\tau}^{t-1}\pp{\bb{\pp{m_t^\dagger M_t}^\dagger S_t}_{t-1}}   & \text{if } \bb{S_t}_{t-1}\notin\text{ISG}\pp{t-1}\quad \& \quad t> \tau,\\
S_t  & \text{if } t=\tau,
\end{cases}
\end{equation} 
\end{widetext}
where $\bb{S_t}_{t-1}$ is the action of replacing the time coordinate of $S_t$ by $t-1$. The output of the ancestry $\text{ance}_{\tau}^{t}\pp{S_t}$ is a time vector with two types of components. The element at time $\tau$ is the part of $S_t$ that was known at time $\tau$, while all the other elements that compose the time vector are contributions to $S_t$ that come after.

The ancestry is important because it describes how the outcome of the measurement of $S_t$ is linked to previous spacetime errors. The phase contribution of an error $E_\tau$ to the measurement of $S_t$ can only depend on measurements made before the time slice $\tau$. Any measurement that comes after will be consistent with the error. Therefore, the phase contribution of such an error on the measurement of a stabilizer $S_t$ depends on the commutation relation between $E_\tau$ and the part of $S_t$ that was known at time $\tau$
\begin{equation} \label{individual_detect}
  \pi_\tau\circ \text{ance}_{\tau}^{t}\pp{S_t},
\end{equation}
where $\pi_\tau \circ \bm{v}$ is the projector on the $\tau$ component of a time vector $\bm{v}$. We can combine the contribution of each time slice $0\leq \tau \leq t-1$ inside a time vector
\begin{equation}\label{eq:detec}
  \bm{\mathcal{D}}_t = \pp{\pi_j\circ \text{ance}_{j}^{t-1}\pp{\bb{S_t}_{t-1}}}_{j=0}^{t-1},
\end{equation}
that we call a detector at time $t$. In more details, a detector at time $t$ corresponds to the measurement of a stabilizer $S_t$ that was already in ISG$\pp{t-1}$. The complete phase contribution coming from a spacetime error $E$ is then given by
\begin{equation}\label{eq:phase_detect}
 s_t = \sum_{j=0}^{t-1} \omega\pp{\mathcal{D}_{t,j}, E_j},
\end{equation}
where $\omega$ is an alternating bilinear form that outputs the phase obtained from the commutation of the two operators. \Cref{eq:phase_detect} is called the syndrome $s_t$ of the spacetime error $E$ by a measurement of $S_t$. For qubit codes, any spacetime error $E$ such that \cref{eq:phase_detect} is $0 \mod{2}$ is undetectable by the detector $\bm{\mathcal{D}}_t$. By combining the detectors from each round of measurement, we can construct a parity matrix $H_{j,k}=\omega\pp{\mathcal{D}_{j,k-1},\cdot}$ for $j\geq k$ and $H_{j,k}=0$ otherwise (we consider that the bilinear form $\omega$ has a matrix representation). \Cref{eq:phase_detect} becomes $\bm{s} = H\bm{E} \mod{2}$, and similarly to the static codes, the kernel modulo $2$ of the parity matrix $H$ gives the set of undetectable spacetime errors. The inference window $\bb{t-\mu\pp{t},t}$ of a dynamical code is the smallest time window such that every element of $\text{ISG}\pp{t}$ is inferred by measurements in the last $\mu\pp{t}$ rounds. In a general dynamical code, the inference window of a given stabilizer $S_t$ might not be fixed and might depend on the time coordinate $t$. A dynamical code for which $\mu\pp{t}$ increases slower than $t$ is called a bounded-inference dynamical code. In many examples, $\mu\pp{t}=\mu$ is a constant, and for static codes $\mu\pp{t}=1$.

After finding a set of generators for the undetectable errors, it remains to find which undetectable errors lead to a logical failure, and which do not. The set of benign errors, introduced in \cite{delfosse_2023} and further used in \cite{blackwell_2025}, induces equivalence classes of logical failure among undetectable errors. Those benign errors are composed of two types of generator: vacuous errors and sandwiching errors. Vacuous errors are the application of a Pauli error $M_{t,j}$, right after $M_{t,j}$ was measured. Such an error applies an unmeasurable global phase to the state and does not influence future measurements. A sandwiching error is one of the form
\begin{equation}
   \bb{E_{t}}_{t+1}^\dagger E_{t},
\end{equation} 
where $E_{t}$ is a Pauli operator that commutes with the measurement $\bm{M}_{t+1}$. This error lies on both time side of a measurement round $\bm{M}_{t+1}$, but since it commutes with it, the part at time $t$ gets annihilated at time $t+1$. Combinations of benign generators cannot lead to a logical fault, and they form the class of acceptable undetectable errors. Two errors that differ by a benign error are defined to be in the same equivalence class, and therefore, correspond to the same logical operator.

The distance of a dynamical code is defined as the minimum weight of an undetectable error that is not benign. To determine the equivalence class of an undetectable error $\bm{E}$ with support on rounds $\bb{t_0,t_1}$ in a bounded-inference dynamical code, it suffices to push the error $\bm{E}$ to the round $t_1+\mu\pp{t_1}+1$ with benign errors. It was shown in Ref. \cite{blackwell_2025} that such a spacetime error $\bm{E}$ is equivalent to a single time error $E_{t_1+\mu\pp{t_1}+1}$ and that its logical class correspond to the single time logical class of $E_{t_1+\mu\pp{t_1}+1}$ inside $\text{ISG}\pp{t_1+\mu\pp{t_1}+1}$.

\subsection{Subsystem codes}
Subsystem codes are a generalization of stabilizer codes where some of the logical qubits, called gauge qubits, are not used to store logical information. The information contained in the logical qubit can be teleported to a gauge qubit, changing the code definition. There is thus a strong connection between the notion of code switching and subsystem codes. In this section, we give a brief description of subsystem codes, and we refer the reader to Ref. \cite{poulin_2005} for further details. 

A subsystem code is defined by a set of gauge operators $\{G_i\}\in \mathcal{P}_n$, which generate the gauge group $\mathcal{G}$. $\mathcal{G}$ is, in general, not an Abelian group. The stabilizer group $\mathcal{S}$ is then the center of the gauge group
\begin{equation}
  \mathcal{S} = Z(\mathcal{G}).
\end{equation}
$\mathcal{G}$ can be decomposed into a set of stabilizer generators and a set of pure gauge generators describing the $X$ and $Z$ operators of multiple gauge qubits. The logical operators are elements of the normalizer of the stabilizer group that are not in the gauge group, $\mathcal{L} = N(\mathcal{S}) \setminus \mathcal{G}$. By projecting all gauge qubits in a chosen Pauli basis, one recovers a standard stabilizer code. Therefore, subsystem codes generalize stabilizer codes. Representatives for the logical operators are separated into two groups: those that act non-trivially on the logical qubits and those that act non-trivially on the gauge qubits. They are called bare logical operators and dressed logical operators, respectively.\\

Subsystem codes can be advantageous because they allow for simpler syndrome extraction circuits and can lead to lower-weight stabilizers compared to traditional stabilizer codes.

\subsection{\label{subsec:Prel_GKP} GKP codes}
In this section, we start by reviewing the construction of multimode GKP codes \cite{gottesman_2001,harrington_2001,conrad_2022,royer_2022} and their connection to lattices. GKP codes are stabilizer codes which utilize the infinite-dimensional Hilbert space of $N$ bosonic modes to encode a finite-dimensional subspace. Each mode is equipped with a creation and annihilation operator $\hat{a}_k$ and $\hat{a}^\dagger_k$ with commutation relation $[\hat{a}_j,\hat{a}^\dagger_k] = \delta_{j,k}$, where $k$ is the mode index. In this work, we will instead prefer to work with the quadrature coordinate operators $\hat{q}_k = (\hat{a}_k+\hat{a}^\dagger_k)/\sqrt{2}$ and $\hat{p}_k = -i(\hat{a}_k-\hat{a}^\dagger_k)/\sqrt{2}$ with $\hbar=1$. When arranged in a vector ${\hat{\bm{\xi}}} = (\hat{\bm{q}},\hat{\bm{p}})$ under the $qqpp$ convention, the commutation relation becomes $[\hat{\xi}_j,\hat{\xi}_k] = i \Omega_{jk}$ for $j,k=1,...,2N$, where the matrix $\Omega$ is defined as
\begin{equation}
  \Omega_{2N} =  \begin{pmatrix}
    0 & 1 \\
 -1 & 0 
    \end{pmatrix} \otimes I_N.
\end{equation}
Unitaries generated by linear Hamiltonians in the quadrature operators 
\begin{equation}\label{eq:trans_def}
 \hat{T}\left(\bm{v}\right) = {\rm e}^{-il \bm{v} \Omega\hat{\bm{\xi}}^T },
\end{equation}
act as translations of the quadratures $\hat{\bm{\xi}} \to\hat{\bm{\xi}} +l\bm{v}$ for $\bm{v}\in\mathbb{R}^{2N}$. Those unitaries obey the commutation relations
\begin{subequations}
  \begin{equation}\label{eq:trans_comm}
 \hat{T}\pp{\bm{u}}\hat{T}\pp{\bm{v}} = {\rm e}^{-i l^2\bm{u}\Omega\bm{v}^T}\text{ } \hat{T}\pp{\bm{v}}\hat{T}
 \pp{\bm{u}} ,
\end{equation}
\begin{equation}\label{eq:trans_comb}
 \hat{T}\pp{\bm{u}}\hat{T}\pp{\bm{v}} ={\rm e}^{\frac{-i l^2}{2}\bm{u}\Omega\bm{v}^T}\text{ } \hat{T}\pp{\bm{u}+\bm{v}},
\end{equation}
\end{subequations}
where we can define the symplectic form 
\begin{equation}
  \omega\pp{\bm{u},\bm{v}} = \bm{u}\Omega\bm{v}^T.
\end{equation}
When using translations with units of $l=\sqrt{2\pi}$, the commutativity of two translation operators is  
tantamount to the symplectic form being an integer:
\begin{equation}\label{commute_condition}
 [\hat{T}\pp{\bm{u}},\hat{T}\pp{\bm{v}}] = 0 \iff \omega\pp{
 \bm{u},\bm{v}} \in \mathbb{Z}. 
\end{equation}
For GKP codes, translation operators are used to construct finite-dimensional representations of the Weyl-Heisenberg group on a finite-dimensional logical subspace. This embedding is achieved by measuring a commuting set of $n\leq 2N$ translation operators $\left\{\hat{T}\pp{\bm{s}_j} \right\}$ with linearly independent vectors $\left\{\bm{s}_j\right\}\in\mathbb{R}^{2N}$, which form the generating basis for the abelian stabilizer group
\begin{equation}\label{eq:GKP_stab}
 \mathcal{S} = \left< \mathrm{e}^{-2\pi i \varphi_1}\hat{T}(\mathbf{s}_1),...,\mathrm{e}^{-2\pi i \varphi_n}\hat{T}(\mathbf{s}_{n})\right>.
\end{equation}
The matrix $S$ with row vectors $\left\{\bm{s}_j\right\}$, complemented with a phase vector $\varphi\pp{S} = \pp{\varphi_1,...,\varphi_n}$, uniquely determine the stabilizer group $\mathcal{S}$. Usually, it is not possible to measure a non-self-adjoint operator like $\hat{T}\pp{\bm{s}_j}$. However, the operators
\begin{subequations}
\begin{equation}\label{eq:trans_cos}
  \cos\pp{l\bm{s}_j \Omega\hat{\bm{\xi}}^T } = \frac{\hat{T}\pp{\bm{s}_j}^\dagger+\hat{T}\pp{\bm{s}_j}}{2}
\end{equation}
and
\begin{equation}\label{eq:trans_sin}
  \sin\pp{l\bm{s}_j \Omega\hat{\bm{\xi}}^T } = \frac{\hat{T}\pp{\bm{s}_j}^\dagger-\hat{T}\pp{\bm{s}_j}}{2i}
\end{equation}
\end{subequations}
are self-adjoint operators, and they can be combined for an indirect measurement of $\hat{T}\pp{\bm{s}_j}$ through the relation
\begin{equation}
  \hat{T}\pp{\bm{s}_j} = \cos\pp{l\bm{s}_j \Omega\hat{\bm{\xi}}^T }-i\sin\pp{l\bm{s}_j \Omega\hat{\bm{\xi}}^T }.
\end{equation}
When the operators of \cref{eq:trans_cos,eq:trans_sin} are measured with outcome $\cos\pp{2\pi \varphi_j}$ and $\sin\pp{2\pi \varphi_j}$ respectively, a measurement of $\hat{T}\pp{\bm{s}_j}$ would give the eigenvalue $\mathrm{e}^{2\pi i \varphi_j}$. Alternatively, we will also write that $\varphi\pp{s_j} = \varphi_j$ is the outcome of a measurement of $\hat{T}\pp{\bm{s}_j}$. The measurement of both \cref{eq:trans_cos,eq:trans_sin} is also equivalent to a measurement of the modular quadrature operator 
\begin{equation}
  l\bm{s}_j \Omega\hat{\bm{\xi}}^T \mod 2\pi.
\end{equation}
To be able to measure simultaneously all translations in \cref{eq:GKP_stab}, the translations operators need to commute. The abelianity condition of $\mathcal{S}$ translate to the skew-symmetric matrix
\begin{equation}\label{eq:A_mat}
 A = S\Omega S^T
\end{equation} 
being integral. The matrix $A$ is called the symplectic Gram matrix. When $A$ is integral, the GKP stabilizer group of \cref{eq:GKP_stab} is isomorphic to its label group $\Lambda$ that contain all integer combinations of the row vectors of $S$
\begin{equation}\label{eq:stab}
  \Lambda = \{\bm{a}S\,| \, \mathbf{a} \in \mathbb{Z}^{n}\},
\end{equation}
which form a lattice. The set of generators $S$ of the lattice group $\Lambda$ is not unique. From a given generator matrix $S$, all equivalent generator matrices $S^\prime$ are related to $S$ by an invertible matrix of integer
\begin{equation}
  S^\prime = RS \quad R\in \text{GL}\pp{n,\mathbb{Z}}.
\end{equation}
Such a change of basis changes the symplectic gram matrix to 
\begin{equation}
  A^\prime = RAR^T.
\end{equation}
It is always possible to find a matrix $R_D\in \text{GL}\pp{n,\mathbb{Z}}$ that brings $A$ into the canonical form:
\begin{equation}\label{eq:canonical}
  A_D = R_DAR_D^T = \pp{\Omega_2 \otimes D} \oplus \bm{0}_{n-2k},
\end{equation}
with $D$ a diagonal matrix of integers greater than $0$ called the type of the lattice. The corresponding generator matrix $S_D = R_DS$ is called a symplectic basis of $\Lambda$. Unless stated otherwise, we work with such a basis for the remainder of the paper. Under a change of basis $S^\prime = RS$, the phase vector becomes
\begin{equation}\label{eq:unimod_phase_upd}
  \varphi\pp{RS}^T = R\varphi\pp{S}^T + \vec{\text{diag}}\pp{\frac{RA_{\text{\lltriangle}}R^T}{2}}\mod{1},
\end{equation}
where $A_\text{\lltriangle}$ is the lower triangular part of the symplectic Gram matrix $A$.

We remark that the definition of the GKP stabilizer group of \cref{eq:GKP_stab} deviates from the standard definition proposed in \cite{conrad_2022,royer_2022}, as we enable the set $\left\{\bm{s}_j\right\}\in\mathbb{R}^{2N}$ to not be a complete basis of $\mathbb{R}^{2N}$. This modification enables the description of codes that are not fully stabilized yet or codes that include auxiliary degree of freedom. This also means that $\text{span}\pp{S}$, together with the form $\omega$, might not be a symplectic subspace, which is needed to embed logical qudits. The canonical form of the matrix $A$ in \cref{eq:canonical} naturally splits the generator of $S$ in two subsets. The span of the first $2k$ rows of $S$ $\pp{S_{\text{symp}}}$ generates a symplectic subspace, while the last $n-2k$ rows of $S$ $\pp{S_{\text{iso}}}$ are skew-orthogonal ($\omega\pp{\bm{s}_j,\bm{s}_k}=0\,\text{if}\, j,k\geq n-2k$) and span what we call an isotropic subspace. The full lattice is therefore generated by two sublattices
\begin{equation}
  \Lambda = \Lambda_{\text{symp}}+\Lambda_{\text{iso}},
\end{equation}
with each sublattice generated by their subset of generators
\begin{equation}
 \Lambda_{\text{symp}} = \{\bm{a}S_{\text{symp}}\,| \, \mathbf{a} \in \mathbb{Z}^{2k}\},
\end{equation}
and
\begin{equation}
  \Lambda_{\text{iso}} = \{\bm{a}S_{\text{iso}}\,| \, \mathbf{a} \in \mathbb{Z}^{n-2k}\}.
\end{equation}
Similarly to the qubit case, the logical Pauli operators are the operators that commute with the set of stabilizers $\pp{\mathcal{N}(\mathcal{S})/\mathcal{S}}$, with the difference that we restrict the mapping to the symplectic subspace of $\Lambda$. In other words, the set of translations that are logical operators have labels in 
\begin{equation}
  \Lambda^* = \Lambda_{\text{symp}}^*+\Lambda_{\text{iso}},
\end{equation}
where 
\begin{equation}\label{eq:lambda_star}
  \Lambda_{\text{symp}}^{*} = \{\mathbf{{u}} \in\text{span}\pp{S_{\text{symp}}}\, | \, \omega\pp{\bm{u},S_{\text{symp}}} \in \mathbb{Z}^{2k}\}.
\end{equation}
Notice here that the dual operation is only applied to the symplectic subspace. The lattice $\Lambda^*$ is also called the symplectic dual lattice of $\Lambda$. In general, we use a superscript $*$ on a lattice $\Lambda$ to indicate that we take the symplectic dual of the symplectic subspace only. A generator matrix for the symplectic dual lattice $\Lambda^*$ is 
\begin{equation}
 S^{*} = \pp{A_{\text{symp}}^{-1} \oplus I_{\text{iso}}} S, 
\end{equation}
where 
\begin{equation}
  A_{\text{symp}}^{-1} = \Omega_2 \otimes D^{-1}
\end{equation}
is the inverse of the symplectic gram matrix, again restricted to the symplectic subspace. Another description for the set $\Lambda^*$ is 
\begin{equation}\label{eq:def_lambda_star}
  \Lambda^* = \left\{ \pp{\bm{a}A^{-1} +\bm{b}}S \text{ } | \text{ }  \bm{a},\bm{b}\in\mathbb{Z}^{n}  \right\},
\end{equation}
with $A^{-1}$ the Moore-Penrose inverse of the matrix $A$ defined in \cref{eq:A_mat}. In the case of a full-rank lattice, $n=2k = 2N$, the definitions above are identical to those found in the literature \cite{harrington_2001,royer_2022,conrad_2022}. A consequence of the definitions of \cref{eq:stab,eq:def_lambda_star} is that 
\begin{equation}\label{eq:hierarchy_lattice}
\Lambda \subseteq \Lambda^*. 
\end{equation}
Each logical Pauli operators is then associated with one of the $\text{det}(\Lambda) \coloneq \text{det}(A_{\text{symp}})$ cosets of $\Lambda^*/\Lambda$. For each of the $k$ non-zero components $D_{jj}$ in the decomposition of \cref{eq:canonical}, a qudit of dimension $D_{jj}$ is defined and the total dimension of the encoded Hilbert space $\mathcal{H}_D = \mathbb{C}^{D_{11}}\otimes...\otimes\mathbb{C}^{D_{kk}}$ is $\text{Det}\pp{D}$. The dimension of $\mathcal{H}_D$ is also connected to the matrix $A$ by the relation
\begin{equation}
  \text{Det}_+\pp{A} = \text{Det}\pp{D^2},
\end{equation}
where $\text{Det}_+$ is the product of positive eigenvalues.
Each encoded qudit of dimension $D_{jj}$ can also be understood as a combination of multiple qudits of smaller dimension. Each prime factor $p_l$ in the prime decomposition 
\begin{equation}\label{eq:prime_decomp}
  D_{jj} = \prod_{l\in\mathbb{N}} p_l^{a_l}
\end{equation}
of $D_{jj}$ gives rise to an encoded qudit of dimension $p_l^{a_l}$ (See Ref. \cite{bermejo-vega_2014}), with corresponding qudit Pauli operators $\hat{T}\pp{\bm{s}_j/p_l^{a_l}}$ and $\hat{T}\pp{\bm{s}_{j+k}/p_l^{a_l}}$. Finally, the distance of a GKP code, denoted $d\pp{\Lambda}$, is defined as the Euclidean norm of the shortest non-trivial logical vector in $\Lambda^*$:
\begin{equation}
  d\pp{\Lambda} = \min_{\bm{\lambda}^*\in\Lambda^*\setminus \Lambda} |\bm{\lambda}^*|.
\end{equation}

\section{\label{sec:SubGKP} Subsystem GKP codes}
\noindent Qudit subsystem quantum codes are qudit quantum codes in which not all the encoded space is used. An unused logical qudit, called a gauge qudit, can be modified and measured without affecting the logical information contained in the rest of the encoded qudits \cite{kribs_2005}. A main advantage of subsystem codes is that the additional degree of freedom over the gauge qudits sometimes enables a stabilization protocol with the measurement of lower weight operators. While the advantage usually comes at the cost of a lower encoding rate and reduced error protection, the performance of subsystem codes, like the Bacon-Shor code \cite{bacon_2006}, remains interesting. In the GKP setting, reducing the length of the vectors being measured is particularly relevant because, in most applications on GKP, the fidelity of the measurement increases when the vectors are shorter or involve less quantum modes. 

In this section, we define an equivalent construction of subsystem codes for the GKP setting, generalizing the lattice picture of \cref{subsec:Prel_GKP} to include gauge degrees of freedom. We find that the measurement of translation operators outside the dual lattice $\Lambda^*$ of a GKP code cannot be described within the stabilizer formalism. Such a measurement partially reveals information on the gauge qudits that is not accounted for by the static code alone. We find that scaled GKP codes are naturally equipped to describe those gauge degrees, and we use them to define subsystem GKP codes. The generalization of GKP codes to subsystem GKP codes is also the essential building block to be able to introduce the dynamical GKP codes in the subsequent sections. 

\subsection{\label{subsec:Sca_GKP}Scaled GKP codes}
\noindent Starting from a lattice $\Lambda$ with generator matrix $S$ such that its symplectic Gram matrix $A$ is in its canonical form, consider the (non-uniformly) scaled GKP code obtained from the lattice $\Lambda_B$ with generator matrix $BS$, where $B$ is a diagonal matrix of positive integers. The action of the matrix $B$ on $S$ is to scale each generating vector $\bm{s}_j$ by an integer $B_{jj}$. By doing so, $\Lambda_B$ can be understood as $\Lambda$ with the elements which are not products of $B$ removed.  Therefore, $B_{jj}\bm{s}_j\in \Lambda$, from which we get the relation
\begin{equation}\label{eq:hierarchy_stab}
  \Lambda_B \subseteq \Lambda,
\end{equation}
between both stabilizer lattices. The scaling of the generator matrix $S$ has also an effect on the symplectic Gram matrix, which is transformed to 
\begin{equation}\label{eq:init_scale_gram}
  A_B = BAB^T.
\end{equation}
Considering that $A$ is in the canonical form of \cref{eq:canonical}, any scaling of the $n-2k$ last vectors cannot affect the size of the encoded space. Therefore, we fix the scaling of the last $n-k$ vectors to $1$, and we write the matrix $B$ as
\begin{equation}\label{eq:scale_decomposition}
  B = B_q \oplus B_p \oplus I_{n-2k},
\end{equation}
where $B_q, B_p\in\text{diag}\pp{\mathbb{N}^k}$. This decomposition enables us to evaluate \cref{eq:init_scale_gram} to
\begin{equation}\label{eq:scale_gram}
  A_B = \pp{\Omega_2\otimes B_qB_p D} \oplus \bm{0}_{n-2k},
\end{equation}
where the effect of the scaling on the type of the lattice can be identified as
\begin{equation}\label{eq:lattice_type_trans}
  D\to G D,
\end{equation}
where $G=B_qB_p$. The scaling increases the dimension of the logical space, in which we can potentially encode more information. To be able to encode additional qudits in the encoded subspace of dimension $D_{jj}$, based on \cref{eq:prime_decomp,eq:lattice_type_trans}, the element $G_{jj}$ needs to contain additional prime factors that are not contained in $D_{jj}$. If this condition is met by the scaling $G_{jj}$, any operation acting only on those additional encoded qudits cannot affect the initial information encoded in $\Lambda$ and the additional qudits can be considered as gauge qudits. Hence, scaled GKP codes with additional prime factors are the equivalent of qudit subsystem codes, and a GKP code can always be viewed as a scaled GKP code with its gauge qudits measured. We can formalize this result by comparing the lattice that describes the initial GKP code and its scaled version. Coming back to the scaling $S\to BS$, the generator of the symplectic dual lattice is transformed to
\begin{equation}
  S_B^* =  B^{-1}S^*.
\end{equation}
With the generators of the scaled lattice and its dual, we can write both sets as
\begin{equation}
  \Lambda_B = \{\mathbf{a}BS\text{ } | \text{ } \mathbf{a} \in \mathbb{Z}^{n}\},
\end{equation} 
and
\begin{equation}\label{eq:scale_dual}
  \Lambda_B^* = \left\{ \pp{\bm{a}B^{-1}A^{-1} +\bm{b}}S \text{ } | \text{ }  \bm{a},\bm{b}\in\mathbb{Z}^{n}  \right\},
\end{equation}
respectively. Usually, a factor of $B$ would multiply $\bm{b}$, but by fixing the scaling to $1$ in the non-symplectic subspace in \cref{eq:scale_decomposition}, both definition are equivalent. Comparing \cref{eq:def_lambda_star,eq:scale_dual}, we get
\begin{equation}\label{eq:hierarchy_logical}
  \Lambda^* \subseteq \Lambda_B^*.
\end{equation}
Finally, combining \cref{eq:hierarchy_lattice,eq:hierarchy_stab,eq:hierarchy_logical}, we find the hierarchy 
\begin{equation}\label{eq:hierchy}
    \Lambda_B \subseteq \Lambda \subseteq \Lambda^* \subseteq \Lambda_B^*
\end{equation}
between the original and scaled lattices. This relation means that a GKP code is always contained within a scaled version of itself.

\subsection{Subsystem GKP codes}\label{sec:Sub_GKP}
\noindent In the last subsection, we found that a subset of scaled GKP codes can be understood as subsystem GKP codes. This construction contrasts to the habitual construction of qudit subsystem codes, that are described by gauge groups. In the present subsection, we link both descriptions by showing that when we consider a GKP code $\Lambda$ with symplectic gram matrix in its canonical form and a scaling matrix $B$, it is always possible to define a GKP gauge group $\mathcal{G}$ and its corresponding gauge lattice $\Lambda_g$ that describe the exact same code as $\Lambda_B$. As a result of this connection, we find that all subsystem GKP codes can be understood as scaled GKP codes.

\noindent To arrive at this result, we start by constructing the gauge lattice $\Lambda_g$ from the initial lattice $\Lambda$ and its scaled version $\Lambda_B$. Let $\Lambda$ describes a GKP code encoding qudits of dimension $D$ and $\Lambda_B$ a scaled GKP code of dimension $B_qB_p D = GD$. Each prime factor in $GD$ can represent a different encoded qudit (see \cref{eq:prime_decomp}). Our goal here is to preserve the initial qudit structure of $\Lambda$ inside the scaled lattice. To achieve this, we reorganize the dimension matrices with 
\begin{equation}\label{eq:prime_sep}
  \tilde{G} = \gcd\pp{D, G}^{-1} G \qquad \tilde{D} = \gcd\pp{D, G} D
\end{equation}
where the $\gcd$ is extended element-wise on diagonal matrices and the product $GD=\tilde{G}\tilde{D}$ is preserved. This regrouping of prime factors guarantees that 
\begin{equation}\label{eq:matrix_gcd}
  \gcd\pp{\tilde{D}, \tilde{G}} = I,
\end{equation}
and that the lattice $\Lambda_B$ encodes $k$ logical qudits of dimension $\tilde{D}_{jj}$ and $k$ gauge qudits of dimension $\tilde{G}_{jj}$, based on \cref{eq:prime_decomp}. The initial qudits of dimension $D$ are encoded in potentially bigger qudits of dimension $\tilde{D}$, but with compatible prime factors. The encoded subspace can then be factorized in a tensor product
\begin{equation}
  \mathcal{H} = \mathcal{H}_C \oplus \mathcal{H}_C^\perp =
  \pp{\mathcal{H}_{\tilde{D}}\otimes\mathcal{H}_{\tilde{G}}} \oplus 
  \mathcal{H}_C^\perp,
\end{equation}
where each subspace can further be factorized as a composition of qudits
\begin{equation}
  \mathcal{H}_{\tilde{D}} \otimes \mathcal{H}_{\tilde{G}} \simeq \pp{\bigotimes_{\tilde{D}_{jj}}\mathbb{C}^{\tilde{D}_{jj}}}\otimes \pp{\bigotimes_{\tilde{G}_{jj}}\mathbb{C}^{\tilde{G}_{jj}}}.
\end{equation}
The separation of the prime factors in \cref{eq:prime_sep} is necessary to keep the initial encoded qudits of $\Lambda$ inside the first subsystem $\mathcal{H}_{\tilde{D}}$. While $\Lambda_B^*$ contains all logical Pauli operations, it is useful to consider the subsets $\Lambda_{\tilde{D}}^*$ and $\Lambda_{\tilde{G}}^*$, which act non-trivially only on their respective subsystem. Choosing a symplectic basis $S_B$ for $\Lambda_B$, we can write explicitly the generator matrix for the three mentioned lattices as
\begin{subequations}
  \begin{equation}
    S_{B}^* = \pp{\Omega\otimes \tilde{G}^{-1}\tilde{D}^{-1}}S_{B},
  \end{equation}
  \begin{equation}
    S_{\tilde{D}}^* = \pp{\Omega\otimes \tilde{D}^{-1}}S_{B},
  \end{equation}
  and
  \begin{equation}
    S_{\tilde{G}}^* = \pp{\Omega\otimes \tilde{G}^{-1}}S_{B},
  \end{equation}
\end{subequations}
respectively. In the context of GKP codes, the gauge group $\mathcal{G}$ correspond to the set of all translation operations $\hat{T}\pp{\bm{g}},\bm{g}\in \Lambda_B^*$ that can be decomposed on $\mathcal{H}_{\tilde{D}} \otimes \mathcal{H}_{\tilde{G}}$ with an identity action on $\mathcal{H}_{\tilde{D}}$
\begin{equation}\label{eq:operation_decomposition}
  \hat{T}\pp{\bm{g}}\ket{\overline{D},\overline{G}} = \pp{I_D\otimes \overline{P_G}}\ket{\overline{D},\overline{G}},
\end{equation}
where we identified an arbitrary basis state of $\mathcal{H}_{\tilde{D}} \otimes \mathcal{H}_{\tilde{G}}$ by $\ket{\overline{D},\overline{G}}$, and $\overline{P_G}$ is a logical operator that acts solely on the gauge qudits. Any vector $\bm{g}$ that has the property of \cref{eq:operation_decomposition} is also an element of $\Lambda_{\tilde{G}}^*$ with generator $S_{\tilde{G}}^*$. We will alternatively refer to $\Lambda_{\tilde{G}}^*$ as the gauge lattice $\Lambda_g$, for which we can give an alternative description as 
\begin{subequations}
\begin{equation}
  \Lambda_g = \left\{\bm{\lambda}_B^*\in \Lambda_B^*\,|\,\omega\pp{\bm{\lambda}_B^*, \bm{\lambda}_{\tilde{D}}^*}\in\mathbb{Z}\,\forall \bm{\lambda}_{\tilde{D}}^*\in\Lambda_{\tilde{D}}^* \right\},
\end{equation}
which is the symplectic centralizer of $\Lambda_{\tilde{D}}^*$ inside $\Lambda_B^*$ \footnote{This notation is inspired from the connection of those object at the group level. For the gauge group, we have that $\mathcal{G}$ is the centralizer of $\mathcal{L}_{\tilde{D}}$ in $\mathcal{L}_B$ : $\mathcal{G}=C_{\mathcal{L}_B}\pp{\mathcal{L}}$, where $\mathcal{L}_B$ is the group of all logical operations on the scaled code, while $\mathcal{L}_{\tilde{D}}$ is the group of all logical operation on the first subsystem. We decided to extend the notation to the underlying label groups.}
\begin{equation}\label{eq:connection_gauge_scale}
  \Lambda_g = C_{\Lambda_B^*}\pp{\Lambda_{\tilde{D}}^*}.
\end{equation}
\end{subequations}
The lattice $\Lambda_g$, being isomorphic to $\mathcal{G}/U\pp{1}$, correspond exactly to the gauge group in the context of qudit subsystem codes. This shows that for any scaled GKP code $\Lambda_B$ obtained by scaling the GKP code $\Lambda$, we can construct a unique subsystem GKP code $\Lambda_g$:
\begin{equation} \label{eq:mapping}
  \pp{\Lambda,\Lambda_B} \to \Lambda_{g}.
\end{equation}
In the reverse direction, we find that the lattice $\Lambda_B$ is the set of operations that leaves both the logical and gauge qudits untouched
\begin{equation}
  \Lambda_B = \mathcal{Z}\pp{\Lambda_g} =  C_{\Lambda_g}\pp{\Lambda_g}.
\end{equation}
In other words, it is the symplectic center of $\Lambda_g$. From $\Lambda_g$ and $\Lambda_B$, it is then possible to recover $\Lambda_{\tilde{D}}^*$ with the relation
\begin{equation}
  \Lambda_{\tilde{D}}^* = C_{\Lambda_B^*}\pp{\Lambda_{g}},
\end{equation}
and obtain a unique mapping 
\begin{equation}
  \Lambda_g \to \pp{\Lambda_{\tilde{D}},\Lambda_B}.
\end{equation}
The lattice $\Lambda$ with generator matrix $S$ can be recovered from $\Lambda_g$ if, and only if
\begin{equation}
  \Lambda_{\tilde{D}} = \Lambda,
\end{equation}
which only happen when $\tilde{D} = D$. In such a case, both representations $\Lambda_g$ and $\pp{\Lambda,\Lambda_B}$ are equivalent in the sense that they contain the same information on the code.

\noindent To finish this section on subsystem GKP codes, we introduce important mathematical objects related to the definition of a general subsystem code in the context of GKP codes. The bare lattice is defined as the set of labels for translations that act non-trivially only on the logical qudits
\begin{equation}
  \Lambda_{\text{bare}} = C_{\Lambda_B^*}\pp{\Lambda_{\tilde{G}}^*} = \Lambda_{\tilde{D}}^*.
\end{equation}
Similarly, the dressed lattice is defined as
\begin{equation}
  \Lambda_{\text{dressed}} = \Lambda_B^*\setminus \Lambda_g ,
\end{equation}
and it is the set of labels for translations that can act non-trivially on the logical qudits, but that can also act non-trivially on the gauge qudits.

To complete the construction of subsystem GKP codes, we produce a notion of distance on the subsystem GKP codes. While we could define the distance of the subsystem GKP code to be the smallest logical vector in $\Lambda_B^*$, this would be too restrictive. Some vectors contained in $\Lambda_B^*$ only correspond to action on the gauge qudits, which do not affect the logical information. Similarly, considering only the vectors of $\Lambda_{\text{bare}}$ is not restrictive enough, and it can lead to an overestimation of the distance. A more appropriate definition of distance for a GKP subsystem code is given by the smallest Euclidean distance of a vector in $\Lambda_{\text{dressed}}$ that acts nontrivially on the logical qudits
\begin{equation}
  d\pp{\Lambda_B} = \min_{\bm{\lambda}_B^*\in \Lambda_B^* \setminus \Lambda_g} |\bm{\lambda}_B^*|.
\end{equation}

\section{\label{sec:DynGKP} Dynamical GKP codes}
In analogy to dynamical qubit codes, we define a dynamical GKP code by a time sequence of translation operation measurement. In such a sequence, commuting subsequence are regrouped into rounds of measurement $\left\{\bm{M}_t\right\}$. By measuring translation operators $\bm{M}_{t+1}$ in round $t+1$, the stabilizer group $\mathcal{S}_{t}$ and the logical operators $\mathcal{L}_{t}$ describing the encoding at round $t$ evolve to $\mathcal{S}_{t+1}$ and $\mathcal{L}_{t+1}$. For GKP codes, the modification of the stabilizer group $\mathcal{S}_t$ (logical group $\mathcal{L}_{t}$) is equivalent to a modification of the lattice $\Lambda_t$ ($\Lambda_t^*$). This means that, between the rounds of measurements $\bm{M}_t$ and $\bm{M}_{t+1}$, a dynamical GKP code is described by a static GKP code with stabilizer lattice $\Lambda_t$
\begin{equation}\label{eq:lattice_seq}
\cdots
  \begin{tikzcd}
\Lambda_{t-1} \arrow{r}{\bm{M}_{t}} & \Lambda_{t} \arrow{r}{\bm{M}_{t+1}}
& \Lambda_{t+1}
\end{tikzcd}
\cdots
\end{equation}
In the present section, we discuss the stabilizer update rules that describe the transformation $\Lambda_t\to\Lambda_{t+1}$, the update rules for the logical representatives and the initialization of dynamical GKP codes.

\subsection{Update rules}\label{subsec:UR}
For a GKP code with stabilizer lattice $\Lambda_t$, the set of logical representatives is noted $\hat{T}\pp{\left\{\bm{x}_{t,l}\right\}}$, where each vector $\bm{x}_{t,l}$ is an element of a distinct coset of $\Lambda_t^*$ indexed by $l$. In other words, each logical Pauli operator is associated with a different value of the index $l$. Different representatives for a given coset are related via stabilizer operation
\begin{equation}
  \bm{x}_{t,l}^\prime =  \bm{x}_{t,l} + \bm{\lambda}_t\quad \bm{\lambda}_t\in\Lambda_t.
\end{equation}
Since it is only possible to measure a mutually commuting set of operators, measuring an additional translation operator $\hat{T}\pp{\bm{m}_{t+1}}$ removes all the elements of $\mathcal{S}_t$ that do not commute with the new measurement
\begin{equation}
  \mathcal{S}_{t+1} = \left< C_{\mathcal{S}_t}\pp{\hat{T}\pp{\bm{m}_{t+1}}},\hat{T}\pp{\bm{m}_{t+1}} \right>. 
\end{equation}
At the level of the labels, this relation can also be written with the symplectic commutant as
\begin{equation}\label{eq:lattice_update}
  \Lambda_{t+1} = \left< C_{\Lambda_t}\pp{\bm{m}_{t+1}}, \bm{m}_{t+1}  \right>.
\end{equation}
Therefore, updating the lattice $\Lambda_t$ to $\Lambda_{t+1}$ after an additional measurement of $\hat{T}\pp{\bm{m}_{t+1}}$ is tantamount to finding a basis for 
\begin{equation}
  C_{\Lambda_t}\pp{\bm{m}_{t+1}} = \left\{\bm{\lambda}_t\in \Lambda_t| \omega\pp{\bm{\lambda}_t,\bm{m}_{t+1}}\in \mathbb{Z} \right\}.
\end{equation} 
Such a basis is dependent on the relative nature of the vector $\bm{m}_{t+1}$ with $\Lambda_t$. Before producing the stabilizer update rules, we introduce an equivalence relation, the $y$-equivalence, between vectors of $\mathbb{R}^{2N}$ and each coset of $\Lambda_t$ in $\Lambda_t^*$ indexed by $l$: $\Lambda_{t,l}^*$. The $y$-equivalence relation is important to characterize how a measurement of a translation by $\bm{m}_{t+1}$ influences the code.\\[0.25cm]
\noindent \textbf{Definition :} We write that $\bm{m}_{t+1}$ is $y$-equivalent to $ \bm{x}_{t,l}$
\begin{equation}\label{eq:yequiv}
  \bb{\bm{m}_{t+1}}_{t}\mysim{y} \bm{x}_{t,l},
\end{equation} 
if there exist an integer $y$ such that $y\bb{\bm{m}_{t+1}}_t = \bm{x}_{t,l}$. Because the time index is in general clear from the context, we often drop the square bracket notation and write directly $\bm{m}_{t+1}\mysim{y} \bm{x}_{t,l}$ instead. We also write that $\bm{m}_{t+1}\mysim{y} \Lambda_{t,l}^*$ if $y$ is the smallest integer such that
\begin{equation}\label{eq:yequiv_set}
  \bm{m}_{t+1}\mysim{y} \bm{x}_{t,l} \mod{\Lambda_t}.
\end{equation} 
In this case, we have $y\bm{m}_{t+1} = \bm{x}_{t,l} + \bm{\lambda}_t$ for an arbitrary representative $\bm{x}_{t,l}$ of the coset $\Lambda_{t,l}^*$ and $\bm{\lambda}_{t}\in\Lambda_t$. This definition extend to $\bm{x}_{t,0}\in \Lambda_t$, for which we write $\bm{m}_{t+1}\mysim{y} \Lambda_t$ if $y$ is the smallest integer for which $y\bm{m}_{t+1}\in \Lambda_t$.

\noindent In the mathematical sense, the relation of \cref{eq:yequiv} is not an equivalence relation, but a partial order (reflexive, antisymmetric and transitive). We still refer to \cref{eq:yequiv} as an equivalence relation because if $\bm{m}_{t+1}\mysim{y} \Lambda_{t,l}^*$, measuring the operator $\hat{T}\pp{\bm{m}_{t+1}}$ reveals the information of a logical representative $\hat{T}\pp{\bm{x}_{t,l}}$. Both measurement reveal the same logical information and are, therefore, equivalent. When $\bm{m}_{t+1}$ has non-trivial support on $\mathbb{R}^{2N}\setminus \text{Span}_{\mathbb{Q}}\pp{S}$, we write $\bm{m}_{t+1} \mynsim \Lambda_{t}^*$, where
\begin{equation}
  \text{Span}_{\mathbb{Q}}\pp{S} = \left\{\bm{v}=\sum_l r_l \bm{s}_l | r_l\in\mathbb{Q}  \right\}.
\end{equation}
With this equivalence relation, we can now construct the update rules that consist of three distinct steps: updating the stabilizer lattice generators $S_t\to S_{t+1}$, updating the corresponding phase vector $\bm{\varphi}\pp{S_t}\to\bm{\varphi}\pp{S_{t+1}}$ and updating the logical representatives $\bm{x}_{t,l}\to\bm{x}_{t+1,l}$. For the logical update rules, we omit time indices where no confusion can arise, to simplify the notation.\\[0.25cm]
\noindent \textbf{Stabilizer update rules} :\\[0.25cm]
Let $\ket{\psi}$ be a GKP state codeword stabilized by translation operators with label in the lattice $\Lambda_t$. The lattice $\Lambda_t$ is generated by row vectors $\left\{ \bm{s}_{t,l} \right\}$ of $S_t$ with phase vector $\varphi\pp{S_t}$. Let $\hat{T}\pp{\bm{m}_{t+1}}$ be an additional measurement with outcome $\varphi\pp{\bm{m}_{t+1}}$ (the possible outcomes are described below). To update the lattice generator matrix $S_t$, a generator matrix for $C_{\Lambda_t}\pp{\bm{m}_{t+1}}$ is computed. It is obtained by scaling the generator matrix $S_t$ with a minimal diagonal matrix of integers $Y = \text{diag}\pp{y,1,...,1}$ in a chosen optimal basis $R$, such that only one generator is scaled:
\begin{equation}\label{eq:commutant_upd}
  C_{\Lambda_t}\pp{\bm{m}_{t+1}} = \left<YRS_t\right>.
\end{equation}
The generator matrix of the updated lattice $S_{t+1}$ and the corresponding phase vector $\varphi\pp{S_{t+1}}$ are
\begin{equation}\label{eq:updated_generatro_matrix}
  S_{t+1} = \begin{pmatrix}
YRS_t \\ \hline
\bm{m}_{t+1}
\end{pmatrix},
\end{equation}
and
\begin{equation}
  \varphi\pp{S_{t+1}}^T = \begin{pmatrix}
Y\varphi\pp{RS_t}^T \\ \hline
\varphi\pp{\bm{m}_{t+1}}
\end{pmatrix},
\end{equation}
respectively, where the horizontal bar means that every rows below it has been appended to the matrix. One can further reduce $S_{t+1}$ (with corresponding updates for $\varphi$) with row reduction operations. The explicit derivation for the matrix $Y$ and $R$, together with a proof of the stabilizer update, is presented in \cref{app:stab_upd}.\\[0.25cm]
\noindent \textbf{Measurement outcome} :\\[0.25cm]
The outcome of the new measurement is given by the following rules :
\begin{enumerate}\label{enum}
  \item If $\bm{m}_{t+1}\mysim{y} \Lambda_t$, it is possible to write $y\bm{m}_{t+1}=\bb{\bm{\lambda}_t}_{t+1} = \bb{\bm{a}S_t}_{t+1}\in\bb{\Lambda_t}_{t+1}$ for $\bm{a}\in\mathbb{Z}^{n}$. The phase of $\bm{m}_{t+1}$ is therefore restricted to the set \begin{equation}\label{eq:phase_m}
    \varphi\pp{\bm{m}_{t+1}} = \frac{\bm{a}\varphi\pp{S_t}^T+\frac{1}{2}\bm{a}A_{\text{\lltriangle}t}\bm{a}^T+z}{y}\mod{1}.
  \end{equation}
  At the moment of the measure, $z$ is sampled randomly from $\mathbb{Z}_{y}$.
  \item If $\bm{m}_{t+1}\mynsim\Lambda_t^*$, then $\bm{m}_{t+1} \in \mathbb{R}^{2N}\setminus \text{Span}_{\mathbb{Q}}\pp{\bb{S_t}_{t+1}}$. In this case, the phase $\varphi\pp{\bm{m}_{t+1}}$ cannot be restricted by prior measurements and a random value in $\left[0,1\right)$ is obtained for the measurement.
\end{enumerate}
\noindent \textbf{Logical update rules} :\\[0.25cm]
While we presented the stabilizer update rules and the measurement outcome for both rational and irrational symplectic product between $\bm{m}_{t+1}$ and the rows of $S_t$, the logical update rules are presented here only for the former case. The latter case is handled in \cref{subsec:hom_meas}.

Let $\hat{T}\pp{\bm{x}_k}$ be a logical operator of a GKP code defined with generator matrix $S$ and let $\ket{\psi}$ be a codeword eigenstate of $\hat{T}\pp{\bm{x}_k}$. Following an additional measurement of $\hat{T}\pp{\bm{m}}$ with outcome $\varphi\pp{\bm{m}}$, for each $k$, we update the logical representative $\hat{T}\pp{\bm{x}_k}$ with the following rules.
\begin{enumerate}
  \item If $\omega\pp{\bm{x}_k, \bm{m}}\in \mathbb{Z}$, the logical operator remains unchanged and the state is further stabilized by $\hat{T}\pp{\bm{m}}$. 
  \item If $\bm{m}\mysim{y}\Lambda_{l}^*$ ($l$ and $k$ can be different), then the updated state $\ket{\psi^\prime}$ is an eigenstate of $\hat{T}\pp{\beta_k\bm{x}_k + \bm{\lambda}}$ and $\hat{T}\pp{\bm{m}}$ with outcome $\varphi\pp{\beta_k\bm{x}_k+ \bm{\lambda}}$ and $\varphi\pp{\bm{m}}$, respectively. The scaling $\beta_k$ is the smallest integer such that we can find a stabilizer vector $\bm{\lambda}$ for which $\hat{T}\pp{\beta_k\bm{x}_k + \bm{\lambda}}$ and $\hat{T}\pp{\bm{m}}$ commute, and $\beta_k$ is the denominator of the irreducible fraction $\omega\pp{\bm{x}_k,\bm{x}_l} = \omega\pp{\bm{x}_k,y\bm{m}} = \alpha_k/\beta_k$.
  \item Otherwise, there always exist a minimum $\beta_k\in\mathbb{Z}$ such that we can find $\bm{\lambda}\in\Lambda$ for which $\hat{T}\pp{\beta_k\bm{x}_k + \bm{\lambda}}$ and $\hat{T}\pp{\bm{m}}$ commute. The logical operator is updated to $\hat{T}\pp{\beta_k\bm{x}_k + \bm{\lambda}}$ and complemented with $\hat{T}\pp{\bm{m}}$.
\end{enumerate}
After the update, all logical representatives that are also element of $\Lambda$ are removed. The proof for the logical update rules is presented in \cref{app:Log_rep}. Compared to rule 2, rule 3 also covers the measurement of vectors with a support in the isotropic subspace of the lattice.\\

\noindent If a scaling of $\beta_k>1$ must be applied to $\bm{x}_k$ to find a representative that commutes with the new measurement, the transformation maps both logical states $\ket{\bm{x}_k}$ and $\ket{\beta_k\bm{x}_k}$ to $\ket{\beta_k\bm{x}_k}$. This many-to-one mapping is irreversible, and it means that the logical information encoded using either of those states is unrecoverable. The result is either a measurement of the logical information or an irrecoverable loss of it. In either cases, the size of the encoded Hilbert space is reduced. If instead, we can find a representative $\bm{x}_k^\prime = \bm{x}_k + \lambda$ that commutes with the new measurement for all cosets $\Lambda_k^*$ ($\beta_k=1$), then the mapping is not only one-to-one, but also reversible. This is the case when $\bm{m}\mysim{y}\Lambda$.

\begin{figure*}
  \centering
  \includegraphics[width=\textwidth]{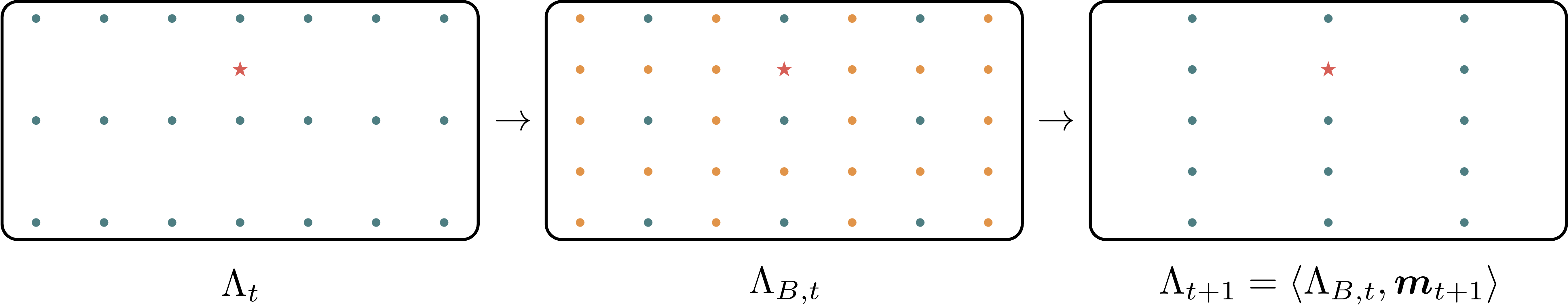}
  \caption{Transformation of the lattice $\Lambda_{t}$ of a qunaught state (left) to the lattice $\Lambda_{t+1}$ (right) due to an additional measurement of $\bm{m}_{t+1}$ (represented by a red star). In each pane, the blue dots represent stabilizer vectors of the corresponding lattice, and the orange dots represent logical vectors. The initial qunaught state of $\Lambda_t$ can be viewed as a logical $\bm{x}$ state of a scaled GKP qubit $\Lambda_{B,t}$. Following the measurement, the code is described by the lattice $\Lambda_{t+1}$ that can be viewed as the $\bm{z}$ logical state of the $\Lambda_{B,t}$ qubit.}
  \label{fig:code_switching}
\end{figure*}

An alternative way to understand the update rules is through subsystem GKP codes. Let us consider an initialized dynamical GKP code at step $t$ and a new measurement of a translation by $\bm{m}_{t+1} \mysim{y} \Lambda_{t,l}^*$, for which either the update rule $1$ or $2$ is applied for all representatives. As presented in \cref{eq:lattice_update}, the updated lattice $\Lambda_{t+1}$ is generated by $C_{\Lambda_t}\pp{\bm{m}_{t+1}}$ and $\bm{m}_{t+1}$, the former is a scaled version of the initial lattice $\Lambda_t$. We can then think of $\Lambda_t$ as the code defined by $\Lambda_{B,t} = C_{\Lambda_t}\pp{\bm{m}_{t+1}}$ after additional logical measurements of the missing generators of $\Lambda_t\setminus \Lambda_{B,t} \subseteq \Lambda_{B,t}^*$. Furthermore, we can also think of $\Lambda_{t+1}$ as the code $\Lambda_{B,t}$, but now with additional logical measurement $\bm{m}_{t+1}\in \Lambda_{B,t}^*$. Therefore, both the initial lattice $\Lambda_t$ and the updated lattice $\Lambda_{t+1}$ can be included in the logical lattice $\Lambda_{B,t}^*$ of a single scaled GKP code $\Lambda_{B,t}$ with
\begin{equation}
  \Lambda_{B,t} = \Lambda_t \cap \Lambda_{t+1}.
\end{equation} 
An example of transformation between $\Lambda_{t}$ and $\Lambda_{t+1}$ is presented in \cref{fig:code_switching}.
The existence of such a scaled lattice $\Lambda_{B,t}$ is that any two lattices from neighboring rounds of measurements $\pp{\Lambda_t,\Lambda_{t+1}}$ form a subsystem GKP code with gauge lattice $\Lambda_{g,t}$, in which the measurement $\bm{m}_{t+1}$ becomes a logical/gauge measurement. The effect of the measurement on the logical information contained in $\Lambda_t$ then depends on the equivalence class of $\bm{m}_{t+1}$ inside $\Lambda_{B,t}^*$. If $\bm{m}_{t+1}$ is in the lattice of dressed vectors, then some logical information is learned, while if it is in the gauge lattice $\Lambda_{g,t}$, then no logical information is learned through the measurement. This alternative vision of GKP codes updates as subsystem GKP codes is useful to gain a deeper understanding of the properties and performances of dynamical GKP codes. 
\subsection{Initialization}
In this section, we consider that all vectors in the measurement round $\bm{M}_{t+1}$ have rational symplectic product with the vectors of $\Lambda_t$. The consequence of this restriction is that $\text{rank}\pp{S_t}$ cannot decrease with additional measurements. Based on the update rules of \cref{eq:updated_generatro_matrix}, $S_t$ can only be scaled when the measurements in $\bm{M}_{t+1}$ are done. Therefore, the $\text{rank}\pp{S_t}$ will necessarily saturate after some finite number of rounds. While the $\text{rank}\pp{S_t}$ might saturate, the size of the encoding might not stabilize at the same time. This needs to be reflected in the initialization time. We will say that a time coordinate $t$ is in the steady stage if the type $D\pp{t}$ of the lattice $\Lambda_t$, defined in \cref{eq:canonical}, and $\text{rank}\pp{S_t}$ are constant for all future time coordinate :
\begin{subequations}
  \begin{equation}
  D\pp{t}=D\pp{t^\prime}\quad \forall t^\prime\geq t,
\end{equation}
and
\end{subequations}
\begin{equation}
 \text{rank}\pp{S_t}=\text{rank}\pp{S_{t^\prime}}\quad \forall t^\prime\geq t.
\end{equation} 
The smallest such time coordinate $T$ is called the initialization time of a dynamical GKP code. In a Floquet GKP code, observe that while the rank of $S$ saturates after at most one period of $P$ rounds, the initialization time can be bigger, and grow linearly with the number of modes $N$. This property can be observed by considering the concatenation of the qubit codes of Refs. \cite{fu_2025,blackwell_2025} with any single mode GKP code. In such examples, the $\text{rank}\pp{S}$ saturates in one period, while the type of the lattice is only stabilized after $O\pp{NP}$ rounds.\\

This last example highlights one property that should be present in any useful dynamic GKP codes. In many dynamic GKP codes, which includes Floquet GKP codes, there exist a lattice of persistent stabilizers $\Lambda_{\text{pers}}$ with generator $S_{\text{pers}}$ such that 
\begin{equation}\label{eq:persistent}
  \Lambda_{\text{pers}} \subseteq \Lambda_t \quad \forall t\geq T, 
\end{equation} 
and such that $\text{rank}\pp{S_{\text{pers}}}=\text{rank}\pp{S_{t}}\quad \forall t\geq T$ \footnote{While it is possible to find pathological cases where this is not true, we consider that it is in general a sound assumption that is true in almost all useful examples.}. This persistent lattice fixes the maximum size of the encoded manifold, and it anchors any GKP code with translation in $\mathbb{R}^{2N}$ to an equivalent with translations in $\mathbb{Q}^{2N}$. This is done by rewriting any computation in terms of the persistent generators $S_{\text{pers}}$. Because all measurements commute with the generators of $\Lambda_{\text{pers}}$, we have that 
\begin{equation}
  M_{t,i} \in \Lambda_{\text{pers}}^*,
\end{equation} 
and therefore that measurement $M_{t,i}$ must be a fraction of a persistent stabilizer. In the case of Floquet GKP codes, it takes at most one period of $P$ rounds to fix the persistent lattice.
\section{\label{sec:EC_DGKP} Error Correction in GKP Dynamical Codes}
In the following section, we focus on the performance of dynamical GKP codes under spacetime translation errors. Precisely, we define the notion of detectors and undetectable errors. By leveraging the lattice of persistent stabilizers, we derive an algorithm to find the distance of a dynamical GKP code. We finish the section with an example, the square-hexagonal GKP code, to demonstrate the tools presented in the section. 
\subsection{Syndrome and undetectable errors}\label{subsec:synd_undetect}
A dynamical GKP code protects the logical information against small spacetime translation errors:
\begin{equation}
  \hat{T}\pp{\bm{\xi}}=\pp{...,\hat{T}\pp{\xi_t},...}.
\end{equation} 
Instead of spacetime translation operators, we will consider their spacetime labels
\begin{equation}
  \bm{\xi}=\pp{...,\xi_t,...}, 
\end{equation} 
with natural group operation
\begin{equation}
  \bm{\xi}+\tilde{\bm{\xi}}=\pp{...,\xi_t+\tilde{\xi}_t,...}.
\end{equation}
The space of vectors $\bm{\xi}$ is isomorphic to the space of translations $\hat{T}\pp{\bm{\xi}}$ with all phase factors removed. The weight of a spacetime error $\bm{\xi}$ is the Euclidean distance of the sum of each time slice error
\begin{equation}
  w\pp{\bm{\xi}} = \norm{\sum_{t\geq 0} \bm{\xi}_t}_2.
\end{equation}
A syndrome is the outcome of a set of measurements that reveal the presence and the type of an error, without collapsing the codespace. In the context of dynamical GKP codes, the outcome of a stabilizer generator $\hat{T}\pp{\bm{\lambda}_{t_1}}$ at round $t_1$ of measurements is affected by previous translation errors and rounds of measurements. Precisely, the outcome $\varphi\pp{\bm{\lambda}_{t_1}}$ is the phase of the eigenvalue obtained when $\hat{T}\pp{\bm{\lambda}_{t_1}}$ is applied on a codeword of round $t_1$
\begin{equation}
  \hat{T}\pp{\bm{\lambda}_{t_1}} \pp{\prod_{j=t_0}^{t_1-1} \Pi_{j+1} \hat{T}\pp{\bm{\xi}_j} \Pi_{j}}\ket{\psi_0},
\end{equation}
where 
\begin{equation}\label{eq:proj_step_j}
  \Pi_{t} = \sum_{\bm{m}_{t,j}\in \mathcal{M}_{t}} \mathrm{e}^{-2\pi i \varphi\pp{\bm{m}_{t,j}}}\hat{T}\pp{\bm{m}_{t,j}}
\end{equation}
is the projector onto the subspace stabilized by the measurement of round $t$. Together, the integer linear combinations of the measurements $\left\{\bm{m}_{t,j}\right\}$ of round $t$ form a lattice $\mathcal{M}_{t}$. From the stabilizer lattice update rules, we know that a stabilizer of round $t_1$ can be written as a product of past measurements, which become a sum of the corresponding labels $\bm{\lambda}_{t_1} = \bm{\lambda}_{t_0} + \sum_{t=t_0}^{t_1}\bm{m}_t$, where $t_0$ is the last time that $\varphi\pp{\bm{\lambda}_{t_0}}$ was obtained. Here, $\bm{m}_t$ is an element of the measurement lattice $\mathcal{M}_{t}$, and not necessarily a measurement itself. Combining everything, we show in \cref{app:eigen} that the outcome for the stabilizer as a function of past measurements and errors is
\begin{equation}\label{eq:stab_out}
    \begin{split}
      \varphi\pp{\bm{\lambda}_{t_1}}-&\varphi\pp{\bm{\lambda}_{t_0}} = \,  \sum_{t=t_0+1}^{t_1} \varphi\pp{\bm{m}_{t}} \\
      &+\sum_{t=t_0}^{t_1-1} \sum_{t^\prime=t_0}^{t} \frac{1}{2}\omega\pp{\bm{m}_{t+1},\bm{m}_{t^\prime}} + \omega\pp{\bm{\xi}_{t},\bm{m}_{t^\prime}}.
\end{split}
\end{equation}
The goal of measuring stabilizers is to gain information on the time sequence $\left\{\bm{\xi}_t\right\}$. Examining \cref{eq:stab_out}, we observe that the measurements of $\hat{T}\pp{\bm{\lambda}_{t_0}}$ and $\hat{T}\pp{\bm{\lambda}_{t_1}}$ reveals the information
\begin{equation}\label{eq:prim_detector}
  \sum_{t=t_0}^{t_1-1} \sum_{t^\prime=t_0}^{t} \omega\pp{\bm{\xi}_{t},\bm{m}_{t^\prime}} \mod{1},
\end{equation}
on the errors. If $\pp{t_1-t_0}$ rounds of measurements are necessary to learn the outcome $\varphi\pp{\bm{\lambda}_{t_1}}$, then only the errors that happened during those rounds can affect the result. The assumption of a past measurement at time $t_0$ in the previous equations can be removed by considering $t_0=0$.

This description fits exactly with the notion of ancestry \cite{blackwell_2025}, and \cref{eq:prim_detector} fits the definition of a detector. In what follows, we adapt our demonstration in the language of Ref. \cite{blackwell_2025}. Given a vector $\bm{\lambda}_t\in \Lambda_t$ and the last measurement vector $\bm{m}_{t}$, the ancestry of $\bm{\lambda}_t$ at time $\tau \leq t$ is defined recursively as the time vector
\begin{widetext}
\begin{equation}\label{eq:gkp_ancestry}
  \text{ance}_{\tau}^{t}\pp{\bm{\lambda}_t} = \begin{cases}
0_t + \text{ance}_{\tau}^{t-1}\pp{\bb{\bm{\lambda}_t}_{t-1}}  & \text{if } \bb{\bm{\lambda}_t}_{t-1}\in\Lambda_{t-1}\quad \& \quad t> \tau,\\
\alpha_t \bm{m}_t + \text{ance}_{\tau}^{t-1}\pp{\bb{\bm{\lambda}_t - \alpha_t \bm{m}_t}_{t-1}}   & \text{if } \bb{\bm{\lambda}_t}_{t-1}\notin\Lambda_{t-1}\quad \& \quad t> \tau,\\
\bm{\lambda}_t  & \text{if } t=\tau,
\end{cases}
\end{equation} 
\end{widetext}
with support over $\bb{\tau,t}$. In the second case, we used the $\mathbb{Z}$-decomposition of $\bm{\lambda}_t$ over the generators $\bm{s}_{t-1,j}$ of $\Lambda_{t-1}$ and the measurement at round $t$ :
\begin{equation}\label{eq:z_decomposition}
  \bm{\lambda}_t = \alpha_t\bm{m}_t + \sum_j z_{t-1,j} \bb{\bm{s}_{t-1,j}}_t,
\end{equation}
for some integer $\alpha_t$ and an element of the measurement lattice $\bm{m}_t\in\mathcal{M}_t$. \Cref{eq:z_decomposition} further implies that $\bb{\bm{\lambda}_t - \alpha_t \bm{m}_t}_{t-1}\in \Lambda_{t-1}$. We can now reinterpret \cref{eq:prim_detector} as 
\begin{equation}
    \sum_{j=0}^{t_1-1} \omega\pp{\bm{\xi}_{j},\mathcal{D}_{t_1,j} } \mod{1},
\end{equation}
with the detector at time $t$
\begin{equation}
  \bm{\mathcal{D}}_{t} = \pp{\pi_j\circ \text{ance}_{j}^{t-1}\pp{\bb{\bm{\lambda}_{t}}_{t-1}}}_{j=0}^{t-1},
\end{equation}
which is a time vector with support over $\bb{0,t-1}$. The last two equations are the exact GKP equivalent of the qubit version in \cref{eq:detec,eq:phase_detect}. In this work, we define a detector to be a measurement of a translation by a vector $\bm{\lambda}_{t}$ such that $\bm{\lambda}_{t} \mysim{y} \Lambda_{t-1}$ for some $y\in\mathbb{Z}$, as defined in \cref{eq:yequiv}. This definition, while equivalent to the one given in \cite{blackwell_2025}, is more appropriate to the context of lattice encoding. Continuing the analogy with the qubit case, we define the phase parity matrix as a block matrix
\begin{equation}\label{eq:phase_parity_matrix}
  H_{j,k} = \begin{cases}
  \mathcal{D}_{j,k-1}\Omega & \text{if }j\geq k\\
  0 & \text{otherwise},
  \end{cases}
\end{equation}
and the syndrome of a spacetime translation error $\bm{\xi}$ is
\begin{equation}\label{eq:gkp_synd}
  \bm{s} = H\bm{\xi} \mod{1}.
\end{equation}
Any undetectable spacetime error is one for which $\bm{s}=0$. Since \cref{eq:gkp_synd} is linear in $\bm{\xi}$, all spacetime errors that correspond to a syndrome $\bm{s}$ are given by
\begin{equation}\label{eq:synd_to_error}
  \bm{\xi} = H^{-1}\bm{s} + \ker_{\mathbb{Z}}\pp{H},
\end{equation}
where $\ker_{\mathbb{Z}}\pp{H}$ is the preimage of $\mathbb{Z}$ under the linear mapping $H$ and $H^{-1}$ is the Moore-Penrose inverse of $H$. In the present context, $\ker_{\mathbb{Z}}\pp{H}$ is a lattice and finding the most probable error $\bm{\xi}$ corresponding to a given syndrome $\bm{s}$ can be understood as solving the closest vector problem (CVP) for $H^{-1}\bm{s}$ in $\ker_{\mathbb{Z}}\pp{H}$ \cite{schnorr_1994,agrell_2002,lin_2023a}. To finish the subsection, we present how to efficiently find generators of $\ker_{\mathbb{Z}}\pp{H}$. To do so, we use the persistent lattice generator $S_{\text{pers}}$ of $\Lambda_{\text{pers}}$ defined in \cref{eq:persistent} to rewrite \cref{eq:gkp_synd} with $\bm{s}=\bm{0}$ as
\begin{equation}\label{eq:und_equation}
  \tilde{\mathcal{D}}\,\cdot\, \text{Diag}\pp{S_{\text{pers}}\Omega S_{\text{pers}}^T} \,\cdot\,\tilde{\bm{\xi}} \in \mathbb{Z}^{N_d}
\end{equation}
where now the detectors $\tilde{\mathcal{D}}$ and the spacetime errors $\tilde{\bm{\xi}}$ are written in the basis of $S_{\text{pers}}$, and the detectors $\tilde{\mathcal{D}}_{j,k}$ are vectors of rational numbers (this is guaranteed by the basis being persistent stabilizer vectors). We have defined $\text{Diag}\pp{S_{\text{pers}}\Omega S_{\text{pers}}^T}$ as the block diagonal matrix with $A_{\text{pers}}$ inserted in each block on the diagonal. Using the lowest common denominator of the detector matrix
\begin{equation}
  \tilde{\mathcal{D}_r} = \tilde{\mathcal{D}}/\text{lcd}\pp{\tilde{\mathcal{D}}},
\end{equation}
we find the integer equation
\begin{equation}\label{eq:simp_kernal}
  \tilde{H}_r \,\cdot\, \tilde{\bm{\xi}} \in \text{lcd}\pp{\tilde{\mathcal{D}}}\mathbb{Z}^{N_d}.
\end{equation}
\Cref{eq:simp_kernal} can be solved directly after finding the Smith normal form of $\tilde{H}_r$.
Each eigenvector with non-zero invariant factor in the Smith decomposition returns a generator of a lattice, while any eigenvector with zero invariant factor generates a continuous degree of freedom in $\ker_{\mathbb{Z}}\pp{H}$. 
\subsection{Distance of dynamical GKP codes}\label{subsec:dist}
In \cref{sec:DynGKP}, we encountered two notions of distance for dynamical GKP code. The first one comes from the construction of a dynamical code in terms of a sequence of static GKP codes $\Lambda_t$, as shown in \cref{eq:lattice_seq}. During a given round $t$, the minimum magnitude of an error that can cause a logical error is given by the distance of the static GKP code $d\pp{\Lambda_t}$. Therefore, we know that there exist at least an error with length
\begin{equation}
  d_{\text{Static}} = \min_{t} d\pp{\Lambda_t}
\end{equation}
that can cause a logical error in the dynamical GKP code. The second notion of distance comes from the rounds of code switching. When a GKP code $\Lambda_t$ is modified to $\Lambda_{t+1}$ by a round of measurement $\bm{M}_{t+1}$, a correctable error in code $\Lambda_t$ can be transformed into an uncorrectable error in the code $\Lambda_{t+1}$. To understand the effect of an error through a round of measurement, we build a subsystem GKP code for each pair of neighboring codes $\pp{\Lambda_t,\Lambda_{t+1}}$
\begin{equation}
  \Lambda_{B,t} = C_{\Lambda_t}\pp{\mathcal{M}_{t+1}}.
\end{equation}
Since $\Lambda_{B,t}$ overlap both static codes, we know that there exist at least one error of weight $d\pp{\Lambda_{B,t}}$ that lead to a logical failure following the code switching step. Combining all steps of code switching, we find that the code switching distance of a dynamical GKP code is
\begin{equation}
  d_{\text{CS}} = \min_{t}d\pp{\Lambda_{B,t}}.
\end{equation}
By construction, $d_{\text{CS}} \leq d_{\text{Static}}$ since $\bm{\lambda}_t^* \in \Lambda_{B,t}^*$ for all $t$. Both definitions give higher bounds on the distance of a dynamical GKP code. In general, the distance of a quantum code is the smallest weight of an undetectable error that is part of a non-trivial logical class. The last statement splits the distance finding into two steps. First, we need to find all undetectable errors, and then, we need to be able to assign a logical class to such errors. The set of undetectable errors, as described above, is the preimage of $\mathbb{Z}$ under the linear mapping of the phase parity matrix $H$ :
\begin{equation}\label{eq:integer_ker}
  \ker_{\mathbb{Z}}\pp{H}.
\end{equation}
While \cref{eq:integer_ker} is concise, it does not prescribe how big a time window should be considered for the accumulation of errors (it is a priori infinite). To resolve this problem, we use the notion of inference window. The inference window of a dynamical GKP code is the minimum integer $\mu\pp{t} \in \bb{0, t-1}$ such that every translation by $\bm{\lambda}_t\in\Lambda_t$ has a fixed value inferred from the last $\mu\pp{t}$ rounds of measurements. We call a dynamical code for which $\mu\pp{t}$ grows slower than $t$ is also called a bounded-inference dynamical GKP code. All codes considered in this work are bounded-inference codes. In addition to the inference window, we define an inference window $\mu_{\text{pers}}\pp{t}$ restricted to the persistent lattice defined in \cref{eq:persistent}. While both are equal in many examples, the two can have distinct values when the initialization time for the type of the lattice and the rank of the generator matrix $S$ are not equal. Because we know that we have access to a full logical rank lattice of persistent stabilizers $\Lambda_{\text{pers}}$, an undetectable and non-trivial error over one persistent inference window $\mu_{\text{pers}}$ needs to have at least weight $d\pp{\Lambda_{\text{pers}}}$. This persistent distance is actually a lower bound on the distance $d$ of the dynamical code. Any error smaller than $d$ would be detectable through a single persistent window, and we must have 
\begin{equation}\label{eq:sandwich_of_distance}
  d\pp{\Lambda_{\text{pers}}} \leq d \leq d_0,
\end{equation}
where $d_0$ is $d_\text{CS}$, but it can also be relaxed to $d_\text{static}$. To obtain a limit on the time window to consider, we use the fact that any two errors separated by more than $\mu_{\text{pers}}$ rounds must be correctable on their own. As a result, we can consider only errors that have non-trivial support over a consecutive set of $\mu_{\text{pers}}$ windows, with a weight of at least $d\pp{\Lambda_{\text{pers}}}$ on each persistent window. After $\lceil{d_0/d\pp{\Lambda_{\text{pers}}}}\rceil $ persistent windows, the smallest undetectable error is bigger than $d_0$, and we can end the search. Therefore, we only need to consider spacetime errors with support over the time slices
\begin{equation}\label{eq:distance_support}
  \bb{ T,T+\left\lceil\frac{d_0}{d\pp{\Lambda_{\text{pers}}} } \right\rceil    \mu_{\text{pers}}},
\end{equation}
to compute the distance, where $T$ is the initialization time of the code. Such errors leave a trace on detectors in the window 
\begin{equation}\label{eq:total_time_window}
  \bb{ T+1,T+\left\lceil\frac{d_0}{d\pp{\Lambda_{\text{pers}}} } +1\right\rceil    \mu_{\text{pers}}},
\end{equation}
as any subsequent measurement will be more than one $\mu_{\text{pers}}$ window away and independent. Over the time window of \cref{eq:total_time_window}, solving \cref{eq:integer_ker} to find all undetectable errors smaller than $d_0$ is now a well-defined problem. 

It now remains to classify each undetectable error in their correct logical coset. To do so, we bring a result of Ref. \cite{blackwell_2025} that states that any undetectable error in a bounded-inference dynamical code with support on the time window $\bb{t_0,t_1}$ can be pushed back to a single round at time $t_1 + \mu\pp{t_1}$ with benign generators. This result still holds for dynamical GKP codes, but the benign generators need to be adjusted. We define a vacuous error in a dynamical GKP code to be a translation by a vector $\bm{m}_{t,j}$ right after the measurement of $\bm{M}_{t}$, but before the measurement of $\bm{M}_{t+1}$. A sandwiching error is a translation error by a spacetime vector $\bm{\xi}=\pp{\bm{\xi}_t, -\bb{\bm{\xi}_{t}}_{t+1}}$ such that 
\begin{equation}
  \omega\pp{\bm{\xi}_t, \bm{m}_{t+1,j}}=0\mod{1} \quad\forall \bm{m}_{t+1,j}\in\bm{M}_{t+1}.
\end{equation}
Together, vacuous and sandwiching errors generate all benign errors. The coset associated with a spacetime translation error supported on $\bb{t_0,t_1}$ is the coset of the error in $\Lambda_{t_1 + \mu\pp{t_1}}$ after being pushed with benign generators. To complete the computation of the distance, we need to go in increasing weight of the spacetime errors given by the generators of $\ker_{\mathbb{Z}}\pp{H}$ and push them to the round 
\begin{equation}
  T+\left\lceil\frac{d_0}{d\pp{\Lambda_{\text{pers}}} } +1\right\rceil    \mu_{\text{pers}}.
\end{equation}
\subsection{Square-Hexagonal code}\label{subsec:examples}
\begin{figure*}
  \centering
  \includegraphics[width=\textwidth]{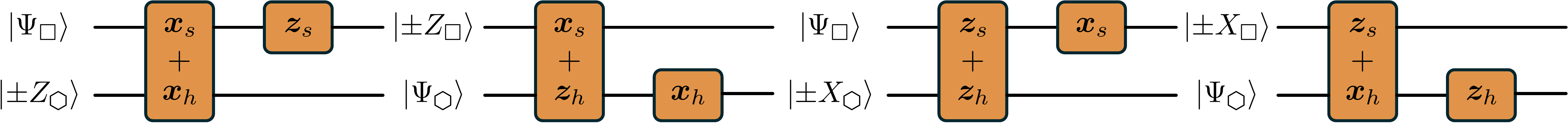}
  \caption{Measurement sequence for the square-hexagonal Floquet code. Each line represent a single mode, while each box correspond to a translation measurement. }
  \label{fig:meas_seq}
\end{figure*}

An example of a dynamical GKP code is a teleportation scheme between a single mode square GKP code and a single mode hexagonal GKP code. While simple, this example illustrates how the tools developed in the previous sections can be used. The square-hexagonal (SH) code is a two-mode period $8$ Floquet GKP code with measurement sequence 
\begin{equation}\label{eq:sh_meas}
  \left\{ \bm{x}_\text{s}+\bm{x}_\text{h}, \bm{z}_\text{s},\bm{x}_\text{s}+\bm{z}_\text{h},\bm{x}_\text{h},\bm{z}_\text{s}+\bm{z}_\text{h},\bm{x}_\text{s}, \bm{z}_\text{s}+\bm{x}_\text{h},\bm{z}_\text{h}  \right\},
\end{equation}
where $\bm{p}_{\text{s}}$ is a vector in the $\bm{p}$ logical coset of the square code, and $\bm{p}_{\text{h}}$ is a vector in the $\bm{p}$ logical coset of the hexagonal code. The measurement sequence is depicted in \cref{fig:meas_seq}.
The square GKP code $\Lambda_{\text{s}}$ has generator matrix $S_{\text{s}} = \sqrt{2}I_2$, while the hexagonal code $\Lambda_{\text{h}}$ has generator matrix
\begin{equation}
  S_{\text{h}} = \frac{1}{\sqrt[\leftroot{-2}\uproot{2}4]{3}} \begin{pmatrix}
  2 & 0\\
  -1 & \sqrt{3}
  \end{pmatrix}.
\end{equation}
\begin{figure}
  \includegraphics[scale = 0.24]{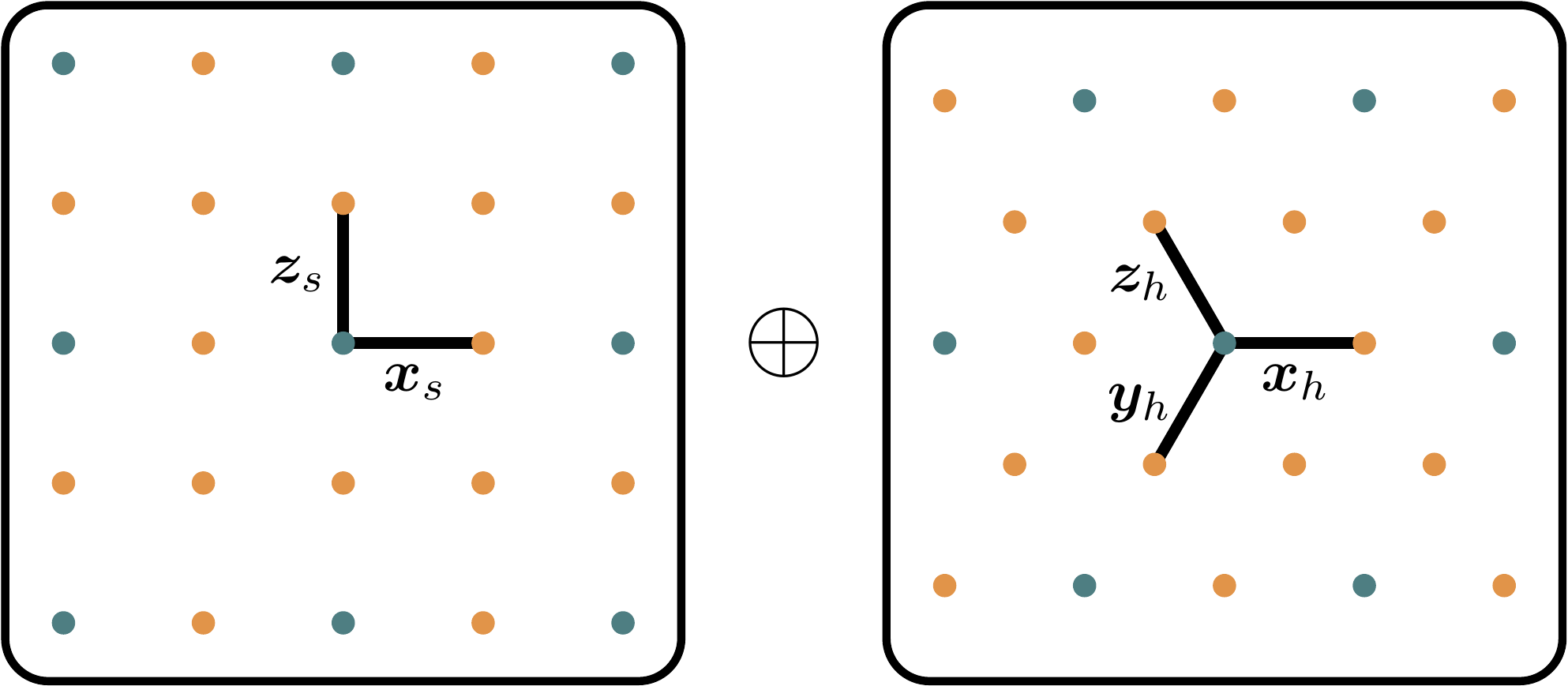}
  \centering
  \caption{The persistent lattice $\Lambda_{\text{pers}}$ of the square-hexagonal GKP Floquet code, represented with blue dots, is the direct sum of the square GKP lattice $\Lambda_s$ (left) with the hexagonal GKP lattice $\Lambda_h$ (right). Every measurement of the sequence in \cref{eq:sh_meas} is an element of $\Lambda_{\text{s}}^* \oplus \Lambda_{\text{h}}^*$, represented by the combination of all blue and orange dots.}
  \label{fig:SH_code}
\end{figure}
The SH code encodes a single logical qubit that is teleported in between the modes every two rounds. Every four rounds, the stabilizers of the maximal persistent lattice
\begin{equation}
  \Lambda_{\text{pers}} = \Lambda_{\text{s}} \oplus \Lambda_{\text{h}},
\end{equation}
are measured, and they correspond to the detectors of \cref{subsec:synd_undetect}. The inference window has width
\begin{equation}
  \mu\pp{t} = \mu_{\text{pers}} = 4.
\end{equation}
Observe that for any measurement $\hat{T}\pp{\bm{m}_t}$ in the sequence with $t\geq T$, we have $\bm{m}_t\mysim{2} \Lambda_{\text{pers}}$. This means that for each measurement $\bm{m}_t$, we have a detector 
\begin{equation}
  \bm{\mathcal{D}}_{t} = \pp{\pi_j\circ \text{ance}_{j}^{t-1}\pp{\bb{2\bm{m}_{t}}_{t-1}}}_{j=0}^{t-1}.
\end{equation}
We construct the syndrome vector from \cref{eq:gkp_synd} with detector support over $\bb{T+1,T+2\mu_{\text{pers}}}$
\begin{widetext}
\begin{equation}\label{eq:syndrome_ex1}
  \bm{s} = 2\begin{NiceArray}{[ccccccc]}
   \bm{z}_\text{s} & 0 & 0 & 0 & 0 & 0 & 0 \\
   \bm{x}_\text{s}+\bm{z}_\text{h} & \bm{x}_\text{s}+\bm{z}_\text{h} & 0 & 0 & 0 & 0 & 0\\ 
   \bm{x}_\text{h} & \bm{x}_\text{h} & \bm{x}_\text{h} & 0 & 0 & 0 & 0\\
   \bm{z}_\text{s}+\bm{z}_\text{h} & \bm{z}_\text{s}+\bm{z}_\text{h} & \bm{z}_\text{s}+\bm{z}_\text{h} & \bm{z}_\text{s}+\bm{z}_\text{h} & 0 & 0 & 0\\
   \bm{x}_\text{s} & \bm{x}_\text{s} & \bm{x}_\text{s} & \bm{x}_\text{s} & \bm{x}_\text{s} & 0 & 0 \\
   \bm{z}_\text{s}+\bm{x}_\text{h} & \bm{z}_\text{s}+\bm{x}_\text{h} & \bm{z}_\text{s}+\bm{x}_\text{h} & \bm{z}_\text{s}+\bm{x}_\text{h} & \bm{z}_\text{s}+\bm{x}_\text{h} & \bm{z}_\text{s}+\bm{x}_\text{h} & 0 \\
   \bm{z}_\text{h} & \bm{z}_\text{h} & \bm{z}_\text{h} & \bm{z}_\text{h} & \bm{z}_\text{h} & \bm{z}_\text{h} & \bm{z}_\text{h}
\CodeAfter
  \UnderBrace[shorten,yshift=1mm]{7-1}{7-1}{T+1}
  \UnderBrace[shorten,yshift=1mm]{7-2}{7-2}{T+2}
  \UnderBrace[shorten,yshift=1mm]{7-3}{7-3}{T+3}
  \UnderBrace[shorten,yshift=1mm]{7-4}{7-4}{T+4}
  \UnderBrace[shorten,yshift=1mm]{7-5}{7-5}{T+5}
  \UnderBrace[shorten,yshift=1mm]{7-6}{7-6}{T+6}
  \UnderBrace[shorten,yshift=1mm]{7-7}{7-7}{T+7}
\end{NiceArray}\cdot \text{Diag}\pp{\Omega}\cdot
\begin{NiceArray}{[c]}
  \bm{\xi}_{T}\\
  \bm{\xi}_{T+1}\\
  \bm{\xi}_{T+2}\\
  \bm{\xi}_{T+3}\\
  0\\
  0\\
  0
  
\end{NiceArray},
\end{equation}
\vspace{0.5cm}
\end{widetext}
where here $\text{Diag}\pp{\Omega}$ is a block diagonal matrix with the antisymmetric form $\Omega$ ($4 \times 4$) on each of its $7$ diagonal block, making the complete size of the matrix $28\times 28$. When we solve \cref{eq:syndrome_ex1}, we find that the only undetectable spacetime errors must be combinations of $\bm{\xi}_t \in \pp{\mathbb{Z}/2}S_{\text{pers}}$. This is equivalent to say that any component of an undetectable error is such that 
\begin{equation}
  \bm{\xi}_t \in  \bb{\Lambda_{\text{pers}}^*}_t = \bb{\Lambda_{\text{s}}^* \oplus \Lambda_{\text{h}}^*}_t.
\end{equation} 
This shows that the distance of the SH Floquet GKP code is
\begin{equation}
  d_{\text{SH}} = d\pp{\Lambda_{\text{pers}}} = d\pp{\Lambda_{\text{s}}} = \frac{1}{\sqrt{2}},
\end{equation}
when considering that a single error in one step can be a logical one with weight $1/\sqrt{2}$.\\

\noindent The full demonstration, as shown here, is sometimes unnecessary. In many cases, we can use \cref{eq:sandwich_of_distance} to reduce, or even pinpoint the actual distance of a dynamical GKP code without any heavy computation. For the SH code, we could have realized that the inequality of \cref{eq:sandwich_of_distance} gives the relation 
\begin{equation}
  \frac{1}{\sqrt{2}} \leq d_{\text{SH}} \leq \frac{1}{\sqrt{2}}.
\end{equation}
On a final note on this example, we identified two main advantages of using the SH code compared to its constituents, the square and the hexagonal code. The first advantage is that using the SH code enables us to use the gate set of both the square and hexagonal codes. Those gate sets include the logical Hadamard gate of the square code with the action
\begin{equation}\label{eq:H_action}
  H_s:\, \bm{x}_s \leftrightarrow \bm{z}_s
\end{equation}
and the logical $HP^\dagger$ gate of the hexagonal code with the action
\begin{equation}\label{eq:HP_action}
  H_hP_h^\dagger:\, \bm{x}_h \rightarrow \bm{z}_h\rightarrow \bm{y}_h\rightarrow \bm{x}_h.
\end{equation}
Those two gates can be easily implemented, both in superconducting (by clock manipulations) and optics (by phase shifters) systems with a Fock $\hat{n}$ Hamiltonian. Fock Hamiltonians correspond to rotation in phase space, which translate to rotation of the lattice. The $\pi/2$ and $2\pi/3$ rotation symmetries of the square and hexagonal lattices, shown in \cref{fig:SH_code}, correspond to the logical operations described in \cref{eq:H_action} and \cref{eq:HP_action}, respectively. Both operations also offer the nice property of being envelope preserving operations, which enhance their fidelity when using finite-energy GKP codes. The second advantage of using the SH code comes from the length of the measured vectors $\bm{m}_t$. Stabilizing and measurement processes have fidelities that are limited by their interaction times to realize an operation, which is in many cases, directly linked to the length of the GKP stabilizers. Using a dynamical GKP code reduces the length of the measurements and should, in theory, improve the fidelity of individual measurement. In the example of the square-hexagonal dynamical GKP code, the longest vector to measure in the sequence has length 
\begin{equation}
  \sqrt{\pp{\frac{1}{\sqrt{2}}}^2 + \pp{\frac{1}{3^{1/4}}}^2} \approx 1.038,
\end{equation}
while the shortest vector to measure in the square GKP code is $\sqrt{2}\approx 1.414$. This is a reduction of approximately $26.6\%$ in the length of the measured vectors. Both the combination of gate sets and reduced measurement lengths are direct dynamical GKP equivalent of the combination of gate set and reduced weight measurements of the dynamical qubit codes. 
\section{Irrational symplectic product measurements}\label{subsec:hom_meas}
In most previous sections, we have considered that our measurement sequence only contains measurements $\bm{m}_t$ such that the measured vectors have rational symplectic product with the stabilizer lattice
\begin{equation}\label{eq:rational_asump}
  \omega\pp{\bb{\bm{m}_t}_{t-1},\bm{\lambda}_{t-1}}\in\mathbb{Q} \,\forall\, \bm{\lambda}_{t-1}\in\Lambda_{t-1}.
\end{equation}
While \cref{eq:rational_asump} is a sound assumption, it leaves behind one of the key difference between continuous and discrete encodings. In the following, we remove this assumption to highlight such differences and find results on the classical simulability of a subset of homodyne measurements in the context of GKP codes. When considering the class of measurements with a non-rational symplectic product with a GKP lattice basis, we are particularly interested in vectors of the type
\begin{equation}\label{eq:scaled_irr}
  \bm{m} = \alpha\bm{\lambda},
\end{equation}
where $\lambda\in\Lambda$ and $\alpha > 0\in \mathbb{R}\setminus\mathbb{Q}$. For the reminder of the section, we drop the time index and focus on a single measurement of the type described in \cref{eq:scaled_irr}. Such measurements are inconsistent with the construction of a symplectic sublattice $\Lambda_B$ inside $\Lambda$ to obtain a subsystem GKP code. Consequently, different update rules must be applied. As we show in the rest of the section, this type of measurement can be equivalent to a homodyne measurement in certain cases.\\

To understand the evolution of a GKP lattice $\Lambda$ under the measurement of \cref{eq:scaled_irr}, we take a detour and focus on finding a more suited basis for the eigenstates of GKP codes. Consider a GKP code with lattice  $\Lambda$ generated by the rows of the matrix $S$, forming a symplectic basis for $\Lambda$. Coming back to \cref{eq:canonical}, the rows of $S$ have the property
\begin{equation}
  \omega\pp{\bm{s}_j,\bm{s}_{N+k}} = D_{jj}\delta_{jk},
\end{equation}
with symplectic invariants $D_{kk}$ of the lattice $\Lambda$. Furthermore, it is possible to show that two vectors $\bm{u}$ and $\bm{v}$ must respect the commutation relation
\begin{equation}
  \bb{\bm{u}\Omega \hat{\bm{\xi}}^T ,\bm{v}\Omega \hat{\bm{\xi}}^T } = i\omega\pp{\bm{u},\bm{v}},
\end{equation}
and so 
\begin{equation}\label{eq:stab_canonical_relation}
  \bb{\bm{s}_j\Omega \hat{\bm{\xi}}^T/D_{jj} ,\bm{s}_k\Omega \hat{\bm{\xi}}^T } = i\Omega_{jk}.
\end{equation}
Backed by the Stone-Von-Neumann theorem, both sets $\left\{\pp{\hat{q}_j,\hat{p}_j}\right\}$ and  $\left\{\pp{\bm{s}_{N+j}\Omega \hat{\bm{\xi}}^T,\bm{s}_j\Omega \hat{\bm{\xi}}^T/D_{jj}}\right\}$ of conjugate variables are equally good to represent the state of the system and their evolution, but the latter is more natural and easier use with general GKP codes. The assertion that both set of conjugate variables are equivalent also stems from the fact that they are related by a symplectic linear transformation. To represent states that respect the restrictions of $\Lambda$, we use the “position” and “momentum” quadrature basis related by a Fourier transform 
\begin{subequations}
  \begin{equation}
    \ket{\bm{s}_{N+j}\Omega \hat{\bm{\xi}}^T = \tilde{q}_j} = \frac{1}{l}\int \text{d}\tilde{p}_j \mathrm{e}^{-i\tilde{q}_j\tilde{p}_j}  \ket{\bm{s}_{j}\Omega \hat{\bm{\xi}}^T/D_{jj} = \tilde{p}_j}
  \end{equation}
  and
  \begin{equation}
    \ket{\bm{s}_{j}\Omega \hat{\bm{\xi}}^T/D_{jj} = \tilde{p}_j} = \frac{1}{l}\int \text{d}\tilde{q}_j \mathrm{e}^{i\tilde{q}_j\tilde{p}_j}  \ket{\bm{s}_{N+j}\Omega \hat{\bm{\xi}}^T = \tilde{q}_j},
  \end{equation}
\end{subequations}
where $l=\sqrt{2\pi}$. For the present purpose, we choose distributions over the “position” basis such that a state can always be written
\begin{equation}
  \ket{\psi} \propto \int \text{d}\tilde{\bm{q}}\,f\pp{\tilde{\bm{q}}} \, \bigotimes_{j=1}^{N} \ket{\bm{s}_{N+j}\Omega \hat{\bm{\xi}}^T = \tilde{q}_j}.
\end{equation}
We also write $S_{\tilde{q}}$ ($S_{\tilde{p}}$) to represent the rectangular matrices with the $N$ first (last) rows of the symplectic basis $S$, such that we also write 
\begin{equation}
  \bigotimes_{j=1}^{N} \ket{\bm{s}_{N+j}\Omega \hat{\bm{\xi}}^T = \tilde{q}_j} = \ket{S_{\tilde{p}}\Omega \hat{\bm{\xi}}^T = \tilde{\bm{q}}^T}.
\end{equation}
Translation operators as defined in \cref{eq:trans_def} have a natural application on such a basis 
\begin{subequations}
  \begin{equation}\label{eq:trans_q_on_basis}
    \begin{split}
    \hat{T}\pp{\bm{\alpha} S_{\tilde{q}}} &\ket{S_{\tilde{p}}\Omega \hat{\bm{\xi}}^T = \bm{\tilde{q}}^T} \\  &\qquad = \ket{S_{\tilde{p}}\Omega \hat{\bm{\xi}}^T = \bm{\tilde{q}}^T + lD\bm{\alpha}^T }
    \end{split}
  \end{equation}
  and 
  \begin{equation}\label{eq:trans_p_on_basis}
    \begin{split}
    \hat{T}\pp{\bm{\beta} S_{\tilde{p}}} &\ket{S_{\tilde{p}}\Omega \hat{\bm{\xi}}^T = \bm{\tilde{q}}^T} \\ &\qquad = \exp\pp{-il\bm{\beta} \bm{\tilde{q}}^T}\ket{S_{\tilde{p}}\Omega \hat{\bm{\xi}}^T = \bm{\tilde{q}}^T}.
    \end{split}
  \end{equation}
\end{subequations}
\Cref{eq:trans_q_on_basis,eq:trans_p_on_basis} enable us to write a general state that inherits the lattice symmetry with generator $S$ and phase vector $\bm{\varphi} = \pp{\bm{\varphi}_{\tilde{q}},\bm{\varphi}_{\tilde{p}}}$ as
\begin{equation}\label{eq:gen_codestate}
  \begin{split}
  \ket{\psi_{S,\bm{\varphi}}} \propto& \sum_{\bm{a}\in\mathbb{Z}^N}\exp\pp{-il^2 \bm{\varphi}_{\tilde{q}}D^{-1}\bb{\bm{a}-\bm{\varphi}_{\tilde{p}}}^T}\\
  & \times g_{D}\pp{l\bb{\bm{a}-\bm{\varphi}_{\tilde{p}}}} \ket{S_{\tilde{p}}\Omega \hat{\bm{\xi}}^T =l\bb{\bm{a}-\bm{\varphi}_{\tilde{p}}}^T },
\end{split}
\end{equation}
where $g_D\pp{\bm{x}}$ is a periodic function that encodes the logical information, and such that 
\begin{equation}
  g_D\pp{\bm{x}+l\bm{a}D} = g_D\pp{\bm{x}}\,\forall\,\bm{a}\in\mathbb{Z}^{N}.
\end{equation}
We are now ready to tackle the main subject of the section. Beginning with a simpler case, consider a qudit of dimension $D$ ($1\times 1$ matrix) encoded in an arbitrary single mode GKP. The lattice that defines the code is generated by a pair of stabilizer vectors $\left\{\bm{s}_1, \bm{s}_2\right\}$ with symplectic product 
\begin{equation}
  \omega\pp{\bm{s}_1, \bm{s}_2} = D.
\end{equation}
Now, if we measure an additional vector $\bm{m} \mynsim \Lambda$ with random outcome of $\varphi_{m}$ such that 
\begin{equation}
  \bm{m} = \alpha \bm{s}_2, \qquad  \alpha \in \mathbb{R}\setminus\mathbb{Q},
\end{equation}
we can update the lattice group using \cref{eq:lattice_update} to find
\begin{equation}
  \Lambda^\prime = \left<\bm{s}_2,\alpha \bm{s}_2\right>. 
\end{equation}
The updated lattice group is in fact not a lattice anymore, and it does not form a symplectic subspace of $\mathbb{R}^2$ that can encode information. The pair $\pp{1,\alpha}$ forms an incommensurable pair. Any real number $\gamma\in\mathbb{R}$ can be approximated with arbitrary precision by an integer combination of those elements because this set is dense in $\mathbb{R}$ \cite{weyl_1910}. Explicitly, we can find $a_1,a_2\in\mathbb{Z}$ such that 
\begin{equation}\label{eq:approx_ineg}
  |\gamma - \pp{a_1 + a_2 \alpha} | \leq \varepsilon
\end{equation} 
for an arbitrary small $\varepsilon$. \Cref{eq:approx_ineg} also means that, for any $\gamma$, the phase of $\hat{T}\pp{\gamma \bm{s}_2}$ can be obtained to any desired precision
\begin{equation}
  \hat{T}\pp{\gamma \bm{s}_2}\ket{\psi^\prime} \approx \exp\pp{2\pi i \bb{a_1\varphi_2+a_2\varphi_m}}\ket{\psi^\prime}.
\end{equation}
The last result includes the measurement of $\hat{T}\pp{\bm{s}_2/D}$ for $\gamma = 1/D$, which means that measuring alpha induces a logical measurement of the code. In summary, in the chosen example, the measurement of $\hat{T}\pp{\alpha \bm{s}_2}$ is a destructive logical measurement. We can further look into the transformation on the codestate of \cref{eq:gen_codestate}. The additional measurement projects the state $\ket{\psi_{S,\bm{\varphi}}}$ into a state that respects the new constraint
\begin{equation}
  \hat{T}\pp{\alpha \bm{s}_2}\ket{\psi_{S,\bm{\varphi}}^\prime} = \exp\pp{2\pi i \varphi_m}\ket{\psi_{S,\bm{\varphi}}^\prime}.
\end{equation} 
For the single mode version of the state in \cref{eq:gen_codestate} to be an eigenvector of $\hat{T}\pp{\alpha \bm{s}_2}$, the term
\begin{equation}\label{eq:irr_const}
  \alpha a \mod 1
\end{equation}
needs to be independent of $a$. The only possible solution is such that $a$ itself has a unique value. The single term of the sum in \cref{eq:gen_codestate} that remains after the measurement is
\begin{equation}\label{eq:quad_proj}
  \ket{\psi_{S,\bm{\varphi}}^\prime} \propto \ket{\bm{s}_{2}\Omega \hat{\bm{\xi}}^T =l\bb{a-\varphi_2}},
\end{equation}
with
\begin{equation}
  \varphi_m \approx -\alpha\pp{a-\varphi_2}\mod{1}.
\end{equation}
We can understand this result as choosing randomly an integer $a$, which is uniquely mapped to a phase $\varphi_m$. As shown in \cref{eq:quad_proj}, the POVM elements of the measurement, after the stabilization of the GKP code, are projectors onto quadrature eigenstates. The translation measurement of $\hat{T}\pp{\alpha \bm{s}_2}$ is therefore equivalent to a homodyne measurement of the quadrature $\bm{s}_2\Omega \bm{\xi}^T$. With the last demonstration, we show a deep connection between the measurement tools used in optic \cite{konno_2024a} and superconducting \cite{sivak_2023a,lachance-quirion_2024a} systems for GKP codes. 

For the more general case of an $N$ modes GKP code, the example of a single mode help us better understand the limitations of the demonstration. It is always possible to find a symplectic basis in which the decomposition of a chosen primitive lattice vector $\bm{\lambda}$ is
\begin{equation}\label{eq:irr_decomp}
  \bm{\lambda} = \bm{b}S_{\tilde{p}},
\end{equation}
where $\bm{b}$ has the minimum possible number $k\leq N$ of non-zero components and $\gcd\pp{\bm{b}} = 1$. The outcome of an irrational measurement of $\hat{T}\pp{\alpha\bm{\lambda}}$ depends directly on the decomposition of \cref{eq:irr_decomp}. If it is possible to decompose $\bm{\lambda}$ into a single symplectic plane over the lattice $\Lambda$ of type $D$ ($k=1$), the single mode case can be applied. When $k\geq 2$ and we force the state in \cref{eq:gen_codestate} to be an eigenstate of $\hat{T}\pp{\alpha\bm{\lambda}}$, we find the condition that
\begin{equation}\label{eq:irr_const_mult}
  \alpha \bm{b}\bm{a}^T \mod 1,
\end{equation}
must be independent of $\bm{a}$. In the following development, we find the solutions of \cref{eq:irr_const_mult} in two steps. We first observe that the only way for \cref{eq:irr_const_mult} to be independent of $\bm{a}$ is to have $\bm{b}\bm{a}^T$ fixed to a single integer $z$. In many cases, fixing $z$ is not enough to fix the vector $a$. In the second step, we solve the equation
\begin{equation}\label{eq:irr_eq}
  \bm{b}\bm{a}^T = z,
\end{equation}
to find the corresponding vectors $\bm{a}$. It is important to note that \cref{eq:irr_eq} has solutions only if $z$ is a multiple of $\gcd\pp{\bm{b}}$, a condition that is always met because $\bm{\lambda}$ is a primitive vector with $\gcd\pp{\bm{b}} = 1$. Removing the $N-k$ trivial components of $\bm{a}$ and $\bm{b}$, we can understand \cref{eq:irr_eq} as a single linear constrain over the lattice $\mathbb{Z}^{k}$, for which the solutions form a $\pp{k-1}$-dimensional lattice. As a result, the projected state is not a single quadrature state, as it was the case when $k=1$, and such a measurement is generally not equivalent to a homodyne measurement. 

To complete the section, we tackle the question of what remains when we consider finite-energy GKP codes instead. To do so, we use the sequence $\left\{p_j/q_j \right\}$ of Diophantine approximations of $\alpha$ that are obtained from the continued-fraction convergents of $\alpha$ \cite{chincin_1997} ($p_j, q_j$ are integers that have no link to the position and momentum variables). As the index goes to infinity $j\to\infty$, the error on the approximation of an irrational number $\alpha$ by the ratios $\left\{p_j/q_j \right\}$ is bounded from above and vanishes
\begin{equation}
  \left| \alpha - \frac{p_j}{q_j}  \right| < \frac{1}{q_{j}^2} \to 0.
\end{equation}
A direct consequence is that the absolute error of distance $e_j$ between a true stabilizer vector $p_j\bm{\lambda}$ and the approximated one $q_j\alpha\bm{\lambda}$ is also bounded from above
\begin{equation}
  |q_j\alpha\bm{\lambda} - p_j\bm{\lambda}| = e_j < \frac{|\bm{\lambda}|}{q_j}.
\end{equation}
As $j\to \infty$, we are guaranteed to have $e_j \to 0$. In the context of finite-energy GKP codes, the measurement of a translation by $q_j\alpha\bm{\lambda}$ would be considered a good approximation of a translation by $p_j\bm{\lambda}$ if 
\begin{equation}\label{eq:approx_cond}
  e_j \leq \Delta,
\end{equation}
where $\Delta$ is the width of each separate peak in phase space \cite{royer_2020a}. The first index $j_{\text{min}}$ for which \cref{eq:approx_cond} is respected correspond to the first good approximation of the measurement. To understand if the measurement is closer to a rational measurement than a homodyne-type one, we compare the length of the approximation $p_{j_\text{min}}|\bm{\lambda}|$ with the size of the envelope $1/\Delta$. If 
\begin{equation}
  p_{j_\text{min}}|\bm{\lambda}| \gg 1/\Delta,
\end{equation}
the stabilizer $p_{j_\text{min}}\bm{\lambda}$ and its approximation $q_{j_\text{min}}\alpha\bm{\lambda}$ meet, for all practical purposes, at infinity. Such a measurement would be very close to a homodyne-type measurement. Otherwise, if the length of the approximation is of the same order or bellow as the size of the envelope 
\begin{equation}
  p_{j_\text{min}}|\bm{\lambda}|  \lesssim 1/\Delta,
\end{equation}
the measurement is a good approximation of a rational measurement that can be understood with the subsystem GKP codes update rules. We observe that the type of the measurement depends mostly on the capacity of approximating the irrational number $\alpha$ with small integers $q_j$ and $p_j$. Therefore, if the goal is to implement a homodyne-type measurement using translation measurement, we need to choose an irrational number $\alpha$ that requires big integer $q_j$ and $p_j$ to approximate it correctly. In the other case the goal is to implement a rational translation measurement, it also shows that we have some protection against small error in the translation measurement. Let say that we measure $\pp{p/q+\varepsilon}\bm{\lambda}$ instead of $\pp{p/q}\bm{\lambda}$ for a small deviation $\varepsilon$. The latter is a good approximation of the former if
\begin{equation}
  q\varepsilon|\bm{\lambda}|  \leq \Delta,
\end{equation}
and both measurements are approximately equivalent. 

\section{\label{sec:Conclusion} Conclusion}
In this work, we have developed a framework for dynamical and subsystem GKP codes, extending the machinery of dynamical qubit codes to the continuous-variable setting of GKP codes. We started by showing that scaled GKP codes, constructed from a base lattice and a scaling matrix, are equivalent to a subsystem construction with explicit bare, dressed, and gauge lattices. This equivalence provides a natural notion of distance for subsystem GKP codes. The identification of subsystem GKP codes and their properties was a necessary step to define correctly dynamical GKP codes as time sequences of translation-operator measurements. We have derived explicit update rules for the different lattices under such sequences of measurements, which allowed us to adapt the notion of ancestry, detectors, and benign errors from the qubit case to the GKP setting. Those connections yielded a prescription to compute the syndrome and the distance of a dynamical GKP code in terms of its persistent stabilizer lattice and its bounded inference window.

We illustrated this framework with the square-hexagonal Floquet GKP code, showing how the distance calculation simplifies considerably once the persistent lattice is identified, and highlights the practical advantages of dynamical GKP codes such as enlarged gate set and reduced measurement vector length. Finally, by relaxing the assumption of rational symplectic products between successive measurements, we established a direct connection between destructive homodyne detection and translation measurements by irrational multiples of a lattice vector, showing that such measurements can be efficiently simulated classically in some cases.

\begin{acknowledgments}
This research was undertaken thanks in part to funding from the Canada First Research
Excellence Fund. This research was also funded by the National Science and Engineering Council with a Canada Graduate Research Scholarship – Doctoral, by the Army Research Office under grant \#W911NF2310045 and by Fonds de de recherche du Québec under grant number 378000.
\end{acknowledgments}

\newpage
\appendix
\section{Lattice generator update}\label{app:stab_upd}
In this section, we derive the transformation rules for the generator matrix $S$ of a lattice $\Lambda$ due to the measurement of the operator $\hat{T}\pp{\bm{m}}$. For the present section, we omit time indices to simplify the notation. We begin with a derivation of \cref{eq:phase_m}. When $y\bm{m}=\bm{\lambda} = \bm{a}S\in\Lambda$ for $\bm{a}\in\mathbb{Z}^{n}$, the phase of the corresponding stabilizer vector is
\begin{equation}
  \varphi\pp{y\bm{m}} = y\varphi\pp{\bm{m}} = \varphi\pp{\bm{a}S} \mod{1}.
\end{equation}
With \cref{eq:unimod_phase_upd}, we have that
\begin{equation}
  y\varphi\pp{\bm{m}} = \bm{a}\varphi\pp{S}^T+\frac{1}{2}\bm{a}A_\text{\lltriangle}\bm{a}^T+z, 
\end{equation}
for a fixed $z\in\mathbb{Z}$, as desired. Next, we update the lattice generator matrix $S$. Before adding $\bm{m}$ as a generator, we scale the generators of $\Lambda$ by minimal integer factor such that all the new generators commute with the additional measurement. To achieve this goal, we compute $\omega\pp{\bm{m},S}$, which allows us separate the generators $\left\{ \bm{s}_l \right\}$ into two sets $V$ and $W$. For $\omega\pp{\bm{m},\bm{s}_l}$ in $\mathbb{Q}$ ($\mathbb{R}\setminus \mathbb{Q}$), we include $\bm{s}_l$ in the set $V$ ($W$). So, the set $V$ contains the generators that have a rational symplectic product with the new measurement, while the set $W$ contains the generators that have an irrational symplectic product with the additional measurement. The set $W$ can be further decompose in multiple sets
\begin{equation}\label{eq:wi_set_def}
  W_i = \left\{\bm{s}_l\in W|\exists \gamma_i\in\mathbb{R}\setminus\mathbb{Q}:\gamma_i\omega\pp{\bm{m},\bm{s}_l}\in\mathbb{Z}\right\},
\end{equation}
each containing generators that have relatively commensurate symplectic product with $\bm{m}$ (i.e. $\omega\pp{\bm{m},\bm{w}_{i,j}}/\omega\pp{\bm{m},\bm{w}_{i,k}}\in\mathbb{Q}\,\forall \bm{w}_{i,j},\bm{w}_{i,k}\in W_i$). Without loss of generality, we can consider that $S$ is ordered in the following way 
\begin{equation}
  S =  \begin{pmatrix}
V \\ \hline 
\left\{W_i\right\} 
\end{pmatrix}.
\end{equation}
The update of $S$ is reduced to the update of each separate set of generators. The update of the sets is done following the rules:
\begin{enumerate}
  \item \textbf{Update of $V$ :} By definition of the set $V$, we write
  \begin{equation}
    \omega\pp{\bm{m},V} = \pp{\frac{\alpha_0}{\beta_0},...,\frac{\alpha_v}{\beta_v}},
  \end{equation}
  with $\alpha_j,\beta_j\in\mathbb{Z}$ forming irreducible fractions. Defining $z_j = \text{Lcm}\pp{\bm{\beta}}\frac{\alpha_j}{\beta_j}\in\mathbb{Z}$, we rewrite
  \begin{equation}\label{eq:almost_V_form}
    \omega\pp{\bm{m},V} = \frac{1}{\text{Lcm}\pp{\bm{\beta}}}\pp{z_0,...,z_k}.
  \end{equation}
  There exists an unimodular transformation $R_V$ that modify \cref{eq:almost_V_form} to
  \begin{equation}
    \omega\pp{\bm{m},R_VV} = \frac{\text{Gcd}\pp{\bm{z}}}{\text{Lcm}\pp{\bm{\beta}}}\pp{1,0,...,0}.
  \end{equation}
  This $R_V$ is found by applying the Hermite normal form algorithm to the vector of integers $\bm{z}$. Denoting by $y$ the denominator of the reduce fraction $\text{Gcd}\pp{\bm{z}}/\text{Lcm}\pp{\bm{\beta}}$, we find that
  \begin{equation}
    V^\prime = \pp{y\oplus I}R_V V,
  \end{equation}
  contains only generators that symplecticaly commute with $\bm{m}$. 
  \item \textbf{Update of $W_i$ :} Using the definition of \cref{eq:wi_set_def}, we can always write
  \begin{equation}
    \omega\pp{\bm{m},W_i} = \gamma_i\pp{z_0,...,z_{w_i}},
  \end{equation}
  with $z_j\in\mathbb{Z}$ and $\text{Gcd}\pp{\bm{z}_{W_i}}=1$. Similarly to the update of $V$, we can find an unimodular transformation $R_{W_i}$ such that
  \begin{equation}
    \omega\pp{\bm{m},R_{W_i}W_i} = \pp{\gamma_i,0,...,0}.
  \end{equation}
  Since no non-zero integer scaling can bring $\gamma_i\in\mathbb{R}\setminus\mathbb{Q}$ into the integers, we obtain that the total transformation of the set $W_i$ is
  \begin{equation}
    W_i^\prime = \pp{0\oplus I}R_{W_i} W_i,
  \end{equation}
\end{enumerate}
with a scaling of $0$ on the first component and of $1$ on all the others. Once we have computed the scaling transformation
\begin{equation}
  Y = \pp{y\oplus I}_V \oplus_i \pp{0\oplus I}_{W_i}
\end{equation}
and the unimodular transformation
\begin{equation}\label{eq:unimod_update}
  R =  R_V \oplus_i R_{W_i},
\end{equation}
we can obtain the updated generator matrix as
\begin{equation}
  S^\prime = \begin{pmatrix}
YRS \\ \hline
\bm{m}
\end{pmatrix}.
\end{equation}
The corresponding phase vector can be computed with \cref{eq:unimod_phase_upd}, and we obtain
\begin{equation}
  \varphi\pp{S^\prime}^T = \begin{pmatrix}
Y\varphi\pp{RS}^T \\ \hline
\varphi\pp{\bm{m}}
\end{pmatrix},
\end{equation}
To finalize the update of the generator matrix, we compute the row echelon form of $S^\prime$ to remove redundant contributions.

\section{Logical representative update}\label{app:Log_rep}
In this section, we derive the minimal scaling that must be applied to a logical representative such that we can find a modified representative that is included  in the updated logical lattice. We will also omit time indices to simplify the notation. Let $\hat{T}\pp{\bm{x}_k}$ be a logical operator of a GKP code defined with generator matrix $S$ and let $\ket{\psi}$ be a codeword eigenstate of $\hat{T}\pp{\bm{x}_k}$. Following an additional measurement of $\hat{T}\pp{\bm{m}}$ with outcome $\varphi\pp{\bm{m}}$, the logical representative $\hat{T}\pp{\bm{x}_k}$ is updated using the following rule :
\begin{quote}
  The logical operator $\hat{T}\pp{\bm{x}_k}$ is updated to $\hat{T}\pp{\beta\bm{x}_k + \bm{\lambda}}$ and complemented with $\hat{T}\pp{\bm{m}}$, where $\beta\in\mathbb{Z}$ is the minimum scaling such that there exist $\bm{\lambda}\in\Lambda$ for which $\omega\pp{\bm{m},\beta\bm{x}_k+\bm{\lambda}}\in\mathbb{Z}$.
\end{quote}
\textbf{Proof :} The goal here is to find the minimal integer $\beta$ for which $\hat{T}\pp{\beta\bm{x}_k + \bm{\lambda}}$ and $\hat{T}\pp{\bm{m}}$ commute. We start by finding the minimum restriction on $\beta$, and we finish by showing that this minimum always leads to a valid solution.\\

The matrix $\tilde{S}=RS$ is a generator matrix of the lattice $\Lambda$ in a new basis, where $R$ is the unimodular matrix in \cref{eq:unimod_update}. A logical representative vector $\bm{x}_k$ is always a fraction of a stabilizer, and it can be written in the new basis as
\begin{equation}
  \bm{x}_k = \sum_{j=1}^{n} \frac{a_j}{b_j}\tilde{\bm{s}}_j, 
\end{equation}
with $a_j,b_j\in\mathbb{Z}$. The problem of updating the representative can be reframed as finding the smallest non-zero integer $\beta$ such that 
\begin{equation}\label{condition_eq}
  \omega\pp{\bm{m},\beta\bm{x}_k+\bm{\lambda}} = \sum_{j=1}^{n} \bb{\beta\frac{a_j}{b_j}+c_j }\omega\pp{\bm{m},\tilde{\bm{s}}_j}\in\mathbb{Z}, 
\end{equation}
where we used that $\bm{\lambda} = \bm{c}S$ for some $\bm{c}\in\mathbb{Z}^{n}$. From the stabilizer update algorithm, we can rewrite this condition as
\begin{equation}\label{eq:big_condition_update}
  \bb{\beta\frac{a_v}{b_v}+c_v }\frac{\text{Gcd}\pp{\bm{z} } }{\text{Lcm}\pp{\bm{\beta} }} +\sum_i \bb{\beta\frac{a_{w_i}}{b_{w_i}}+c_{w_i} }\gamma_i = -u,
\end{equation}
where we used that only one vector in each of the sets $V$ and $\left\{W_i\right\}$ contribute to $\omega\pp{\bm{m},\beta\bm{x}_k+\bm{\lambda}}$ significantly and $u\in\mathbb{Z}$. Each $\gamma_i\in\mathbb{R}\setminus\mathbb{Q}$ contribution cannot be brought in $\mathbb{Z}$ by a rational multiplication. Therefore, each term must cancel itself
\begin{equation}\label{eq:_gamma_part}
  \beta\frac{a_{w_i}}{b_{w_i}}+c_{w_i} =0,
\end{equation}
and $\beta$ must have all $\left\{b_{w_i}\right\}$ as factors. Inserting \cref{eq:_gamma_part} in \cref{eq:big_condition_update}, we find the equation
\begin{equation}\label{eq:number_theo_eq}
  \beta \bb{\frac{\text{Gcd}_r\pp{\bm{z}} a_v}{b_v}} + c_v\bb{\text{Gcd}_r\pp{\bm{z}}} + u\bb{\text{Lcm}_r\pp{\bm{\beta} }} = 0,
\end{equation}
where we have identified with an $r$ subscript the components of the reduced fraction
\begin{equation}\label{eq:frac_red}
  \frac{\text{Gcd}\pp{\bm{z}_V}}{\text{Lcm}\pp{\bm{\beta}_V }} = \frac{\text{Gcd}_r\pp{\bm{z}_V}}{\text{Lcm}_r\pp{\bm{\beta}_V }},
\end{equation}
to obtain $\text{Gcd}\pp{\text{Gcd}_r\pp{\bm{z}_V},\text{Lcm}_r\pp{\bm{\beta}_V }} = 1$. By further reducing the fraction of the first term
\begin{equation}
  \text{Gcd}_r\pp{\bm{z}_V} a_v/b_v = \frac{p}{q},
\end{equation} 
we find that $\beta$ must have at least a $q$ factor for \cref{eq:number_theo_eq} to have a solution. Combining this $q$ factor with conditions obtained from \cref{eq:_gamma_part}, we find the minimum requirement
\begin{equation}\label{eq:beta_cond}
  \beta = \tilde{\beta} \,\text{Lcm}\pp{q,\bm{b}_{w_i}},
\end{equation}
with $\tilde{\beta}\in\mathbb{Z}$. This completes the first part of the proof. It only remains to show that there always exist a valid solution for $\tilde{\beta}=1$. Using \cref{eq:beta_cond}, we must now solve 
\begin{equation}\label{eq:number_theo_eq2}
  \tilde{\beta} \tilde{p} + c_v\bb{\text{Gcd}_r\pp{\bm{z}_V}} + u\bb{\text{Lcm}_r\pp{\bm{\beta}_V }} = 0,
\end{equation}
where 
\begin{equation}
  \tilde{p} = \text{Lcm}\pp{q,\bm{b}_{w_i}} \pp{\frac{\text{Gcd}_r\pp{\bm{z}_V} a_v}{b_v}}. 
\end{equation}
\Cref{eq:number_theo_eq2} is an equation of the type
\begin{equation}\label{eq:general_sol}
  ax+by+cz = 0,
\end{equation}
which can always be solved over $\mathbb{Z}$ for the variables $\pp{x,y,z}$ and constants $\pp{a,b,c}$. The general solution to \cref{eq:general_sol} leads to a solution of \cref{eq:number_theo_eq2} for the variables $\pp{\tilde{\beta},c_v,u}$ and constants $\pp{\tilde{p},\text{Gcd}_r\pp{\bm{z}_V},\text{Lcm}_r\pp{\bm{\beta}_V }}$
\begin{subequations}
  \begin{equation}
  c_v = -\tilde{\beta}\tilde{p}f + k \,\text{Lcm}_r\pp{\bm{\beta}_V },
\end{equation}
\begin{equation}
   u = -\tilde{\beta}\tilde{p}g - k\,\text{Gcd}_r\pp{\bm{z}_V},
\end{equation}
\end{subequations}
with $k\in\mathbb{Z}$ and $f,g$ an arbitrary pair of solution of
\begin{equation}\label{eq:easy_numth_prob2}
  f\bb{\text{Gcd}_r\pp{\bm{z}_V}} + g\bb{\text{Lcm}_r\pp{\bm{\beta}_V }} = 1,
\end{equation}
over the integers. By construction (i.e. \cref{eq:frac_red}) and Bézout theorem \cite{fulton_1969}, there always exist a solution to \cref{eq:easy_numth_prob2} which correspond to a solution of \cref{eq:number_theo_eq} with $\tilde{\beta}=1$. As a result, we can always find an equivalent representative of the logical operator $\hat{T}\pp{\beta\bm{x}_k + \bm{\lambda}}$ that commute with the new measurement of $\hat{T}\pp{\bm{m}}$. \qed\\

To complement this result, we give a more direct proof for the case of rule number 2, when $\bm{m}\mysim{y}\Lambda_{l}^*$. Assuming that there exist $y\in\mathbb{Z}$ such that $y\bm{m}\in\Lambda_l^*$, which is denoted by $\bm{m} \mysim{y}\bm{x}_l$, we write
\begin{equation}
  \bm{m} = \sum_j \frac{a_j}{b_j} \tilde{\bm{s}}_j,
\end{equation}
with $a_j, b_j\in\mathbb{Z}$ and $\tilde{\bm{s}}_j$ the $j$th row out off $K$ of the generating matrix $S$. This decomposition enables us to deduce that 
\begin{equation}
  \omega\pp{\bm{m},S} = \pp{\frac{\alpha_0}{\beta_0},...,\frac{\alpha_K}{\beta_K}},
\end{equation}
again with $\alpha_j, \beta_j\in\mathbb{Z}$. Using the same unimodular matrix $R$ used for the stabilizer update rule, we map $S\to\tilde{S}=RS$ such that 
 \begin{equation}
  \omega\pp{\tilde{S},\bm{m}} = \pp{\frac{\text{Gcd}\pp{\bm{z}}}{\text{Lcm}\pp{\bm{\beta}}},0,...,0}.
\end{equation}
We also denote the reduced fraction with subscript
\begin{equation}\label{eq:numth_red}
  \frac{\text{Gcd}\pp{\bm{z}}}{\text{Lcm}\pp{\bm{\beta}}} = \frac{\text{Gcd}_r\pp{{\bm{z}}}}{\text{Lcm}_r\pp{{\bm{\beta}}}},
\end{equation}
to obtain $\text{Gcd}\pp{\text{Gcd}_r\pp{{\bm{z}}}, \text{Lcm}_r\pp{{\bm{\beta}}}} = 1$. $\Lambda^*$ is defined as the set of vectors that maps $\omega(\tilde{S},\cdot) \to \mathbb{Z}$. Therefore, we find that $y=\text{Lcm}_r\pp{{\bm{\beta}}}$ is the minimal scaling factor such that $y\bm{m}\in\Lambda^*$. We can conclude that 
\begin{equation}\label{eq:logical_symp_prod}
  \omega\pp{\bm{x}_k, \text{Lcm}_r\pp{{\bm{\beta}}} \bm{m}}=\omega\pp{\bm{x}_k, \bm{x}_l} = \frac{\alpha_x}{\beta_x}.
\end{equation}
When we build the condition that $\omega\pp{\bm{m},\beta\bm{x}_k+\bm{\lambda}}\in\mathbb{Z}$ with $\bm{\lambda} = cS$, we find 
\begin{equation}\label{eq:same_eq}
  \beta \omega\pp{\bm{x}_k, \bm{x}_l} + c_0\bb{\text{Gcd}_r\pp{\bm{z}}} + z\bb{\text{Lcm}_r\pp{\bm{\beta} }} = 0,
\end{equation} 
which is the same as \cref{eq:number_theo_eq}. Therefore, we know that it has a solution for $\beta=\beta_x$ from \cref{eq:logical_symp_prod}. \qed\\

\section{Syndrome computation}\label{app:eigen}
In this section, we derive the \cref{eq:stab_out} for the outcome of the stabilizers when updated and affected by errors. Let $\bm{\lambda}_{t_1}$ be a vector for which we obtain the phase $\varphi\pp{\bm{\lambda}_{t_1}}$ after $(t_1-t_0)$ rounds of measurements. The outcome of $\bm{\lambda}_{t_1}$ is related at the same time to the intermediate measurements and the errors during the $(t_1-t_0)$ rounds. To obtain the connection between the stabilizer outcome, the measurements $\bm{m}_t$ at each round and the errors, we compute the eigenvalue for a translation by $\bm{\lambda}_t = \bm{\lambda}_{t_0} + \sum_{t=t_0+1}^{t_1} \bm{m}_t$ in the equation
\begin{equation}\label{eq:big_prod}
  \hat{T}\pp{\bm{\lambda}_{t_1}} \pp{\prod_{t=t_0}^{t_1-1} \Pi_{t+1} \hat{T}\pp{\bm{\xi}_t} \Pi_{t}}\ket{\psi_{t_0}},
\end{equation}
where the projectors on the measured subspace was defined in \cref{eq:proj_step_j}. We will also use $\bm{\lambda}_{t_0} = \bm{m}_{t_0}$ to simplify the notation. Let us focus on a single term of the product
\begin{equation}
  \hat{T}\pp{\bm{\lambda}_t}\Pi_{t} \hat{T}\pp{\bm{\xi}_{t-1}} \Pi_{t-1}.
\end{equation}
Because $\bm{\lambda}_t$ is a stabilizer after round $t$, it commutes with the projector at time $t$
\begin{equation}
  \Pi_{t} \hat{T}\pp{\bm{\lambda}_t}\hat{T}\pp{\bm{\xi}_{t-1}} \Pi_{t-1}.
\end{equation}
Using the properties of the translation operators in \cref{eq:trans_comm,eq:trans_comb}, we arrive at the equivalent equation
\begin{equation}\label{eq:almost_phase}
  \begin{split}
  &\mathrm{e}^{2\pi i \pp{\frac{\omega\pp{\bm{m}_t,\bm{\lambda}_{t-1}}}{2} + \omega\pp{\bm{\xi}_{t-1},\bm{\lambda}_{t-1}}}}\\ &\qquad\times \Pi_{t} \hat{T}\pp{\bm{m}_t}\hat{T}\pp{\bm{\xi}_{t-1}}\Pi_{t-1}\hat{T}\pp{\bm{\lambda}_{t-1}},
  \end{split}
\end{equation} 
when we decompose the last measurement round contribution from the stabilizer. Because $\hat{T}\pp{\bm{m}_t}$ is a stabilizer of the $t$th stabilizer group, we have the relation
\begin{equation}
  \Pi_{t}\hat{T}\pp{\bm{m}_t} = \mathrm{e}^{2\pi i \varphi\pp{\bm{m}_t}}\Pi_{t}.
\end{equation}
Using the last equation in \cref{eq:almost_phase}, we find that the $t$th term in the product of \cref{eq:big_prod} contribute a phase of 
\begin{equation}
  \varphi\pp{\bm{m}_t} + \frac{1}{2}\omega\pp{\bm{m}_t,\bm{\lambda}_{t-1}} + \omega\pp{\bm{\xi}_{t-1},\bm{\lambda}_{t-1}}
\end{equation}
to the total phase. Combining every contribution and rewriting it only in terms of measurements and errors, we arrive at the desired result 
\begin{equation}
    \begin{split}
      \varphi\pp{\bm{\lambda}_{t_1}}-&\varphi\pp{\bm{\lambda}_{t_0}} = \,  \sum_{t=t_0+1}^{t_1} \varphi\pp{\bm{m}_{t}} \\
      &+\sum_{t=t_0}^{t_1-1} \sum_{t^\prime=t_0}^{t} \frac{1}{2}\omega\pp{\bm{m}_{t+1},\bm{m}_{t^\prime}} + \omega\pp{\bm{\xi}_{t},\bm{m}_{t^\prime}}.
\end{split}
\end{equation}
A similar demonstration can be done for the phase update and gauge of the logical representatives.

\end{document}